\documentclass[twocolumn]{aastex631}

\shorttitle{Fundamental Planes for Strong Jet Sources}
\shortauthors{Long et al.}

\usepackage{longtable}
\usepackage{amsmath}
\usepackage{array}
\usepackage{hyperref} 
\usepackage{xcolor}
\usepackage{graphicx}
\usepackage{textcomp}
\newcommand{\angstrom}{\mbox{\normalfont\AA}}

\begin{document}

\title[The Fundamental Planes for the Strong Jet Sources]{Revisiting the Fundamental Planes of Black Hole Activity for Strong Jet Sources}

\correspondingauthor{Ai-Jun. Dong}
\email{aijdong@gznu.edu.cn}
\author[0009-0003-4213-7662]{Qing-Chen. Long}
\affiliation{School of Physics and Electronic Science, Guizhou Normal University, Guiyang 550001, People’s Republic of China; aijdong@gznu.edu.cn}
\affiliation{Guizhou Provincial Key Laboratory of Radio Astronomy and Data Processing, Guizhou Normal University, Guiyang 550001, People’s Republic of China}

\author[0000-0003-4151-0771]{Ai-Jun. $\rm{Dong^{\dagger}}$}
\affiliation{School of Physics and Electronic Science, Guizhou Normal University, Guiyang 550001, People’s Republic of China; aijdong@gznu.edu.cn}
\affiliation{Guizhou Provincial Key Laboratory of Radio Astronomy and Data Processing, Guizhou Normal University, Guiyang 550001, People’s Republic of China}

\author[0000-0001-9389-5197]{Qi-Jun. Zhi}
\affiliation{School of Physics and Electronic Science, Guizhou Normal University, Guiyang 550001, People’s Republic of China; aijdong@gznu.edu.cn}
\affiliation{Guizhou Provincial Key Laboratory of Radio Astronomy and Data Processing, Guizhou Normal University, Guiyang 550001, People’s Republic of China}

\author[0000-0002-9173-4573]{Lun-Hua. Shang}
\affiliation{School of Physics and Electronic Science, Guizhou Normal University, Guiyang 550001, People’s Republic of China; aijdong@gznu.edu.cn}
\affiliation{Guizhou Provincial Key Laboratory of Radio Astronomy and Data Processing, Guizhou Normal University, Guiyang 550001, People’s Republic of China}


\begin{abstract}

Whether the X-ray emissions of strong jet sources originate from disk+coronas or jets is still controversial. In this work, we constructed a strong jet sample containing 50 flat-spectrum radio quasars, 51 low-synchrotron-peaked BL Lac objects and 18 intermediate-synchrotron-peaked BL Lac objects to explore the origin of X-ray emissions. Generally, blazars are the typical radio-loud active galactic nucleus with a powerful jet toward the observer, causing their broadband emissions to be boosted. By considering the Doppler boosting effect, we obtain the intrinsic radio--X-ray correlation and the fundamental plane (FP) of black hole activity for the strong jet sources: the intrinsic radio--X-ray correlation is $L_{\rm{R,int}}\propto L_{\rm{X,int}}^{1.04}$, which favor the jet-dominated mode, the intrinsic FP is $\log L_{\rm{R,int}}=(1.07\pm0.06)\log L_{\rm{X,int}}-(0.22\pm0.10)\log M_{\rm{BH}}-(3.77\pm2.11)$, which can be interpreted by the hybrid mode of jet+standard disk. Our results suggest that the X-ray emissions of strong jet sources are dominated by the jets, but there may also be a small contribution from the disk. In addition, the radio--X-ray correlation and FP of strong jet sources do not have a significant dependence on the Eddington-ratio.
\end{abstract}



\section{Introduction} \label{sec:intro}

Active galactic nucleus (AGNs) are widely believed to be powered by an accreting supermassive black holes \citep[SMBHs; $10^{6-10}\,M_{\odot}$, e.g.,][]{Magorrian1998,Padovani2017}. Understanding the accretion physics and feedback of AGNs has an important implication for insights into the growth and evolution of host galaxies \citep{Blandford2019}. There are two classification of AGNs are observationally distinguished by the radio-loudness parameter, $R=F_{\rm{5GHz}}/F_{\rm{4400}\,\angstrom}$, where $F_{\rm{5\,GHz}}$ and $F_{\rm{4400}\,\angstrom}$ are the rest-frame 5\,GHz radio flux density and optical flux density at $\rm{4400}\,\angstrom$ band, respectively \citep{1989AJ.....98.1195K}. Only $10\%-20\%$ of AGNs are radio-loud AGNs (RL-AGNs) with $R\ge10$, while the rest are radio-quiet AGNs (RQ-AGNs) with $R<10$ \citep{Ivezic2002optical,Kellermann2016radio}. Generally, RL-AGNs exhibit the powerful relativistic jets that are absent in RQ-AGNs \citep{Padovani2017}. 

However, the RL-AGNs/RQ-AGNs dichotomy is often debated in previous studies. For example, (1) this dichotomy has been made along different criteria in previous works, e.g., most of works used $R_{\rm{crit}}=10$ as the criterion of this dichotomy; however, $R_{\rm{crit}}=17$ and $R_{\rm{crit}}=30$ were presented in \cite{Zhang2021radio} and \cite{Bariuan2022FP}, respectively. (2) \cite{Brinkmann2000radio} and \cite{Bonchi2013radio} found that the RL-AGNs/RQ-AGNs dichotomy did not show a bimodal distribution. (3) $R$ depend on the Eddington-ratio \citep[$R\propto 1/\lambda_{\rm{Edd}}$, see][]{Ho2002,Sikora2007ApJ,broderick2011there}, which was interpreted as a change in the accretion mode \citep[$\lambda_{\rm{Edd}}>0.01$ corresponds to a radiatively efficient accretion, while the lower $\lambda_{\rm{Edd}}$ corresponds to a radiatively inefficient advection dominated accretion flow, i.e. ADAF. see][]{Ho2002}, but \cite{Ballo2012} found $R$ is not significantly correlated with $\lambda_{\rm{Edd}}$. 3) Whether the RL-AGNs are the strong jet sources has been controversial \citep{Meyer2011blazar,Keenan2021relativistic}. 

It is generally accepted that the X-ray emissions of RL-AGNs mainly come from jets \citep{yuan2009revisiting,Wang2006black,De-Gasperin2011testing,Bariuan2022FP,Dong2023FP,Wang2024FP}, but there are some views think the X-ray emissions of RL-AGNs come from the disk or corona \citep{Li2008black,li2018black,li2021origin,Zhu2020x}. The fundamental plane (FP) of black hole activity can provide us with an opportunity to gain the insight into the accretion mode of various accreting BHs, as well as to constrain the origin of X-ray emissions \citep[see the works of FP;][]{Merloni2003,Falcke2004,yuan2005radio,yuan2009revisiting,Kording06RefiningA&A,Wang2006black,Li2008black,Gultekin2009FP,De-Gasperin2011testing,Plotkin12,dong2014new,dong2015revisit,Fan2016FP,Xie2017FP,li2018black,Liao2020x,Bariuan2022FP,Wang2024FP,Zhang2024FP}. Over the past 20 yr, the researches on FP have been well developed and refined, such as the FP is sensitive to the weight of adopted sample \citep{Kording06RefiningA&A,Plotkin12}, the FP depends on the radio-loudness \citep{Wang2006black,Li2008black,Bariuan2022FP,Wang2024FP} and the Eddington-ratio of X-ray luminosity \citep[$\lambda_{\rm{Edd}}=\log(L_{\rm{X}}/L_{\rm{Edd}})$, hereafter $\lambda_{\rm{Edd}}$: see][]{yuan2005radio,yuan2009revisiting,dong2015revisit,Xie2017FP,Wang2024FP}. However, the previous works that study the dependence of FP on $\lambda_{\rm{Edd}}$ only focus on the low-luminosity AGNs (LLAGNs); such studies are absent in the brighter blazars.

So far, there have been two different views on the FP and X-ray emissions' origin of RL-AGNs. The prevailing view thought that the FP of RL-AGNs is steeper than of RQ-AGNs, which is attributed to the fact that RL-AGNs are jet-dominated sources \citep[see works of radio-loud sample, e.g.,][]{yuan2009revisiting,Wang2006black,Li2008black,De-Gasperin2011testing,Plotkin12,Xie2017FP,Liao2020x,dong_2021,Dong2023FP,Bariuan2022FP,Wang2024FP}. However, an opposing view was proposed by \cite{li2018black}, who constructed a radio-loud sample that contains $13$ low-excitation radio galaxies (LERGs) from $3CRR$ catalog to study FP. They found a shallower radio--X-ray correlation ($L_{\rm{R}}\propto L_{\rm{X}}^{0.63}$) and FP ($\xi_{\rm{X}}= 0.52$, $\xi_{\rm{M}}= 0.84$), which are consistent with the ADAF mode, and they argued that the X-emissions come from the accretion flows rather than jets. In addition, \cite{li2021origin} firstly found a significant relationship between the mid-infrared to X-ray spectral index $\alpha_{\rm{IX}}$ and the Eddington-ratio $\lambda_{\rm{IR}}$ ($\alpha_{\rm{IX}}-\lambda_{\rm{IR}}$; see their Figure\,2) for a radio-loud sample, which is flatter than the $\alpha_{\rm{OX}}-\lambda_{\rm{O}}$ relationship in RQ-AGNs \citep{Lusso2010x}, implied the disk origin of the X-ray emissions. A similar view has also been claimed in \cite{Zhu2020x}, they found that the jet-linked component is only important for a fraction($<10\%$) of flat-spectrum radio quasars (FSRQs), and the corona-linked component dominates the X-ray emissions of most RL-AGNs.

In the unified model of RL-AGNs, blazars are the typical RL-AGNs that present a powerful jet toward our line of sight\citep{Urry1995}, which causes their broadband emissions to be boosted. Blazars can be separated into two subclasses based on the equivalent width (EW) of the optical emission lines: FSRQs exhibit the strong emission lines ($\rm{EW}>5\,\angstrom$), while BL Lac objects have very weak or no emission lines \citep[$\rm{EW}<5\,\angstrom$; e.g.][]{Scarpa1997high}. In addition, the spectral energy distribution (SED) of blazars shows a double-hump structure, which are characterized by a synchrotron component at lower frequency hump and inverse Compton component at higher frequency hump \citep{Fossati1998unifying,Donato2001hard,Wu2007}. Based on the location of synchrotron peak frequency ($\log \nu_{\rm{peak}}$) in the SED, blazars are classified as low-synchrotron-peaked blazar(LSP; LBL for BL Lac objects), intermediate-synchrotron-peaked blazar(ISP; IBL for BL Lac objects) and high-synchrotron-peaked blazar(HSP; HBL for BL Lac objects). A reliable classification \citep[see][]{Fan2016spectral} is given as the following: for LSP, $\log \nu_{\rm{peak}}<14$; for ISP, $14<\log \nu_{\rm{peak}}<15.3$, and for HSP $\log \nu_{\rm{peak}}>15.3$. It is worth noting that FSRQs have almost no HSP \citep{Fan2016spectral,Ajello2022fourth,Yang2022spectral}. For the blazars' sequence, \cite{Meyer2011blazar} and \cite{Keenan2021relativistic} have suggested that HBLs are the weak jet source, while the rest are the strong jet sources.

In this work, we compiled a strong jet sample from blazars' sequence to reexplore FP and X-ray emissions' origin of strong jet sources. This will help us resolve the possible controversies mentioned above and provide further insights into black hole accretion, jet formation, and feedback. Previous works have suggested that the strong jet blazars have a non-negligible Doppler beaming effect \citep{Wu2007,Nieppola_2008,Yang2022beaming}. Therefore, the Doppler beaming effect should be taken into account for performing physical inference for mechanism of radiation. In addition, the studies of FP depending on $\lambda_{\rm{Edd}}$ are absent in the brighter blazars, we also examine whether the $\lambda_{\rm{Edd}}$ have a significant impact on FP of blazars. The structure of the paper is as follows. In \S\,\ref{Sample}, we detailedly describe our strong jet samples. In \S\,\ref{sec:METHODS AND RESULTS}, we introduce data analysis and the fitting methods and present our best-fit results. The discussions and summaries of our results are presented in \S\,\ref{Discussion} and \S\,\ref{lastpage}, respectively. For this paper, the cosmological parameters are selected as $H_{0} = 70\rm\,km\,s^{-1}\,Mpc^{-1}$, $\Omega_{\Lambda} =0.73$, and $\Omega_{\rm{M}} = 0.27$ \citep{dong2015revisit}.
\section{Sample} \label{Sample}
\cite{Padovani_1995} found that LBLs and HBLs occupy two completely different regions in the $\alpha_{\rm{RO}}-\alpha_{\rm{OX}}$ plane (see their Figure\,12). In addition, the FSRQs and IBLs tend to inhabit in the similar region as LBLs rather than HBLs in the $\alpha_{\rm{RO}}-\alpha_{\rm{OX}}$ plane \citep[detailedly see Figure\,16, Figure\,1, Figure\,10, Figure\,10 and Figure\,6 in][respectively]{Donato2001hard,Padovani2003,Nieppola2006spectral,Fan2016spectral,Yang2022spectral}, which indicates that the energy distribution of FSRQs, IBLs, and LBLs are closer to each other, while the energy distribution of HBLs is very different from them. Moreover, there are some evidences suggesting that the FSRQs, LBLs, and IBLs are strong jet sources that exhibit the radiatively efficient accretion, while the HBLs are weak jet sources that associated with ADAF mode \citep[][because the dichotomy between the strong jet sources and the weak jet sources does not belong to the scope of the discussion in this work, we refer the readers to the above paper for details of this classification]{Meyer2011blazar,Keenan2021relativistic}. Besides, \cite{Donato2005six} found that FSRQs and LBLs tracked the jet-dominated radio--X-ray correlation ($L_{\rm{R}}\propto L_{\rm{X}}^{1.06}$), while the HBLs have a shallower radio–X-ray correlation ($L_{\rm{R}}\propto L_{\rm{X}}^{0.64}$: see their Figure\,4), which agree with ADAF mode in \cite{Merloni2003}. Observationally, the very long baseline interferometry (VLBI) jets of HBLs appear in a plumelike morphology beyond a few milliarcseconds from the core, in contrast to the powerful jets of strong jet blazars that may remain well collimated to large distances \citep{Piner2008parsec}. Therefore, we excluded the HBLs because we only focus on the strong jet sources.

To obtain the largest possible sample of strong jet source, we cross-referenced the FSRQs' catalogue from \citep{Hovatta2009doppler}, BL Lac objects' catalog from \cite{Ye2021unification} and \cite{Wu2014some} with the synchrotron peak frequency $\log \nu_{\rm{peak}}$ of \cite{Ajello2022fourth} and \cite{Xiong2015basic}. Besides, we also add five LBLs that were not used in their works, their 5\,GHz core radio flux densities ($F_{\rm{C,5GHz}}$) taken from NASA/IPAC Extragalactic Database (NED)\footnote{https://ned.ipac.caltech.edu/} or \cite{Yuan2012evolution}, their 5\,GHz Doppler factors $\delta$ are estimated by using Equation\,(6) in \cite{Ye2021unification}. In total, we have collected $50$ FSRQs, $51$ LBLs, and $18$ IBLs for our strong jet sample; they have the available core radio flux density, X-ray flux density, BH mass, $\log \nu_{\rm{peak}}$ and Doppler factor $\delta$. It is noting that we focus only on the strong jet sources. Therefore, the radio-loudness $R$ of blazars are calculated to ensure that they are all RL-AGNs, where the boundary of the radio-loudness dichotomy of $\log R=1.23$ \citep[ref Figure\,6 in][]{Zhang2021radio} is adopted in this work. In order to obtain the values of radio-loudness, we searched the optical flux densities at $\rm4400\,\angstrom$ band in NED and SIMBAD Astronomical Database - CDS\footnote{https://simbad.u-strasbg.fr/simbad/}. For the sources without $F_{\rm4400\,\angstrom}$, the other optical bands are extrapolated to $\rm4400\,\angstrom$ assuming a spectral index of $\alpha_{o}=-0.5$ \citep[$F\propto \nu^\alpha$, see][]{Bariuan2022FP}.

\begin{figure*}
    \centering
    \includegraphics[width=\textwidth]{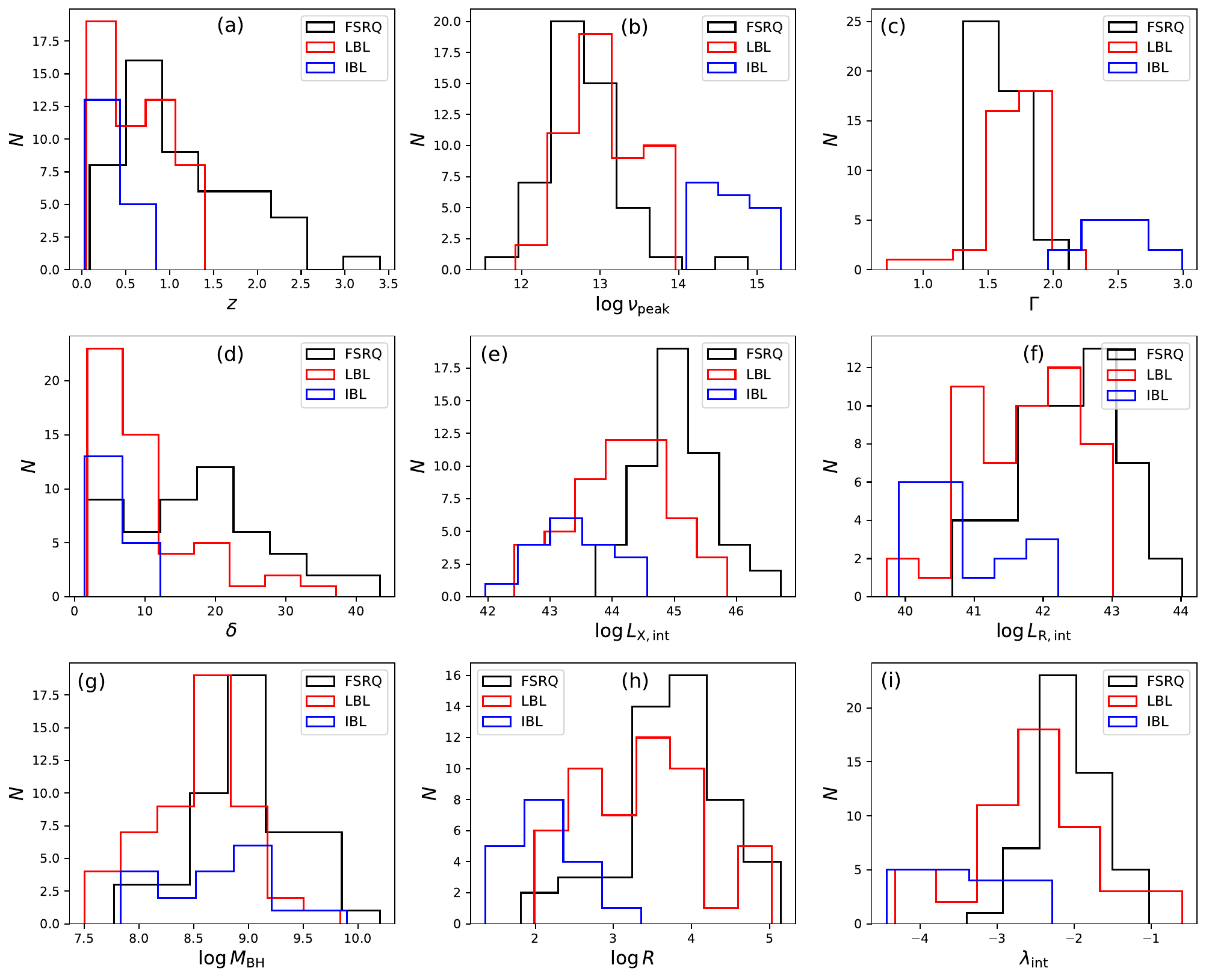}
    \caption{Distributions of the physical parameters for our samples, the black box, red box, and blue box represent FSRQs, LBLs, and IBLs, respectively. (a) Redshift: $z$, (b) Logarithm of the synchrotron peak frequency: log$\nu_{\rm{peak}}$, (c) The power law photon index: $\Gamma$, (d) The 5\,GHz Doppler factor: $\delta$, (e) Logarithm of the intrinsic X-ray luminosity at 2--10\,keV band: log$L_{\rm{X,int}}$, (f) Logarithm of the intrinsic 5\,GHz core radio luminosity: log$L_{\rm{R,int}}$, (g) Logarithm of the black hole mass: log$M_{\rm{BH}}$, (h) Logarithm of the radio-loudness: $\log R$, (i) Eddington-ratio: $\lambda_{\rm{int}}=\log (L_{\rm{X,int}}/L_{\rm{Edd}})$.}
    \label{fig:histogramsources}
\end{figure*}
\subsection{The 5 GHz Core Radio Luminosity}
In order to avoid the contamination from the star formation, in this paper, we use the core 5\,GHz radio flux densities to calculate the 5\,GHz core radio luminosities. For FSRQs, the $F_{\rm{C,5GHz}}$ were taken from NED or CDS. Because the strong jet blazars have the nonnegligible Doppler beaming effect, we taken the Doppler foctor $\delta$ at 22\,GHz and 37\,GHz from \cite{Hovatta2009doppler}, which are extrapolated to the 5\,GHz $\delta$ by using the empirical Equation\,(\ref{Eq3}) (detailedly see \S\,\ref{sec:FP}). For LBLs and IBLs, their $F_{\rm{C,5GHz}}$ mainly come from \cite{Ye2021unification} and \cite{Wu2014some}, the 5\,GHz $\delta$ of their work are also adopted. But once the 5\,GHz $\delta$ are taken from \cite{Ye2021unification}, the $\delta$ values that are estimated by using $q=3+\alpha(\alpha=0)$ are adopted in this work (see their Equation\,(6)). The intrinsic radio luminosities are estimated by taking $\delta$ into account (detailedly see \S\,\ref{sec:FP}). 
\subsection{The 2--10 keV X-Ray Luminosity}
In this paper, the 2--10\,keV X-ray Luminosities are calculated by using 2-10\,keV X-ray flux densities ($F_{\rm{X,2-10}}$). For our blazars, the X-ray observations are mainly detected by \textit{Swift}, \textit{Chandra} and \textit{XMM-Newton}, their X-ray flux densities are taken from NED and the existing literatures. In order to minimize the observational errors, we preferred the X-ray data from \textit{Chandra/XMM-Newton} if the source has the X-ray flux density that was detected by \textit{Chandra/XMM-Newton}, owning to the \textit{Chandra/XMM-Newton} have a higher spatial resolution. In addition, in order to avoid the errors caused by converting through another approximate wave band, we preferred the $F_{\rm{X,2-10}}$. For the sources without the available $F_{\rm{X,2-10}}$, the other approximate wavebands (e.g., 0.5--7\,keV, 0.3--8\,keV, 0.3--10\,keV) are extrapolated to $F_{\rm{2-10\,keV}}$ by using the available power law photon index $\Gamma$ ($F_{\nu}\propto \nu^{1-\Gamma}$):

\begin{equation}
    F_{\rm{X,2-10}}=F_{\rm{X,a-b}}\frac{\int_2^{10}\nu^{1-\Gamma}d\nu}{\int_a^b\nu^{1-\Gamma}d\nu}
    \label{Eq1}
\end{equation}
To reduce the errors caused by the Equation\,(\ref{Eq1}) and (\ref{Eq2}) ($q=2+\alpha_{\rm{X}}=1+\Gamma$), the X-ray flux density and the power-law photon index $\Gamma$ are taken from the same literature. For the sources without the simultaneous power-law photon index $\Gamma$, we adopt the average power-law photon index $\langle\Gamma\rangle$ that are averaged from the sources with the available power-law photon index: $\langle\Gamma\rangle=1.58$ for FSRQs, $\langle\Gamma\rangle=1.71$ for LBLs, $\langle\Gamma\rangle=2.44$ for IBLs. Similarly, the intrinsic X-ray luminosities are estimated by using Equation\,(\ref{Eq2}).

\subsection{The BH Mass}
The BH mass ($M_{\rm{BH}}$) of our blazars are searched in the existing literatures, and their BH mass have multiple estimates. The traditional virial BH mass is also known as the dynamical mass, which adopts an empirical relationship between the broad-line-region (BLR) size and the ionizing luminosity, as well as the measured broad-line width, assuming that the BLR clouds are gravitationally bound by the central BH. This method is usually applied to estimate the BH mass for FSRQs \citep[e.g.,][]{Shen2011catalog,Shaw2012}. Because BL Lac objects have no or weak emission line, their BH mass are usually estimated from the properties of their host galaxies, e.g., $M-\sigma$ and $M-L_{\rm{bulge}}$ relations, where the $\sigma$ and $L_{\rm{bulge}}$ are the stellar velocity dispersion and the bulge luminosity of the host galaxies, respectively. But there are some BL Lacs show the well-detected broad lines; their BH mass therefore are estimated by using the traditional virial method. The prioritization of the choice of BH mass is as follows: $M_{\rm{virial}}>M-\sigma>M-L_{\rm{bulge}}$. $0749+540$ and $0642+449$ do not have the dynamical mass available in the previous works, but we found the absolute magnitude $M_{\rm{R}}=-24.8$ \citep{Maselli2010} for $0749+540$, then we used the $M-M_{\rm{R}}$ relation from \cite{Wu2002supermassive} to estimate its $M_{\rm{BH}}$ ($\log M_{\rm{BH}}=8.94\,M_{\odot}$). The BH mass of $0642+449$ is $\log M_{\rm{BH}}=9.12\,M_{\odot}$ that is obtained by combining the spectroscopic data from \cite{Torrealba2012optical} with Equation\,(2) of \cite{Shen2011catalog}. Finally, our basic data are listed in Table.\,\ref{table2} and Figure\,\ref{fig:histogramsources} shows some parameters' distribution of our sample.

\section{DATA ANALYSIS AND RESULTS}\label{sec:METHODS AND RESULTS}
\begin{figure}
    \vspace{0.3cm}
    \centering{\includegraphics[scale=0.33]{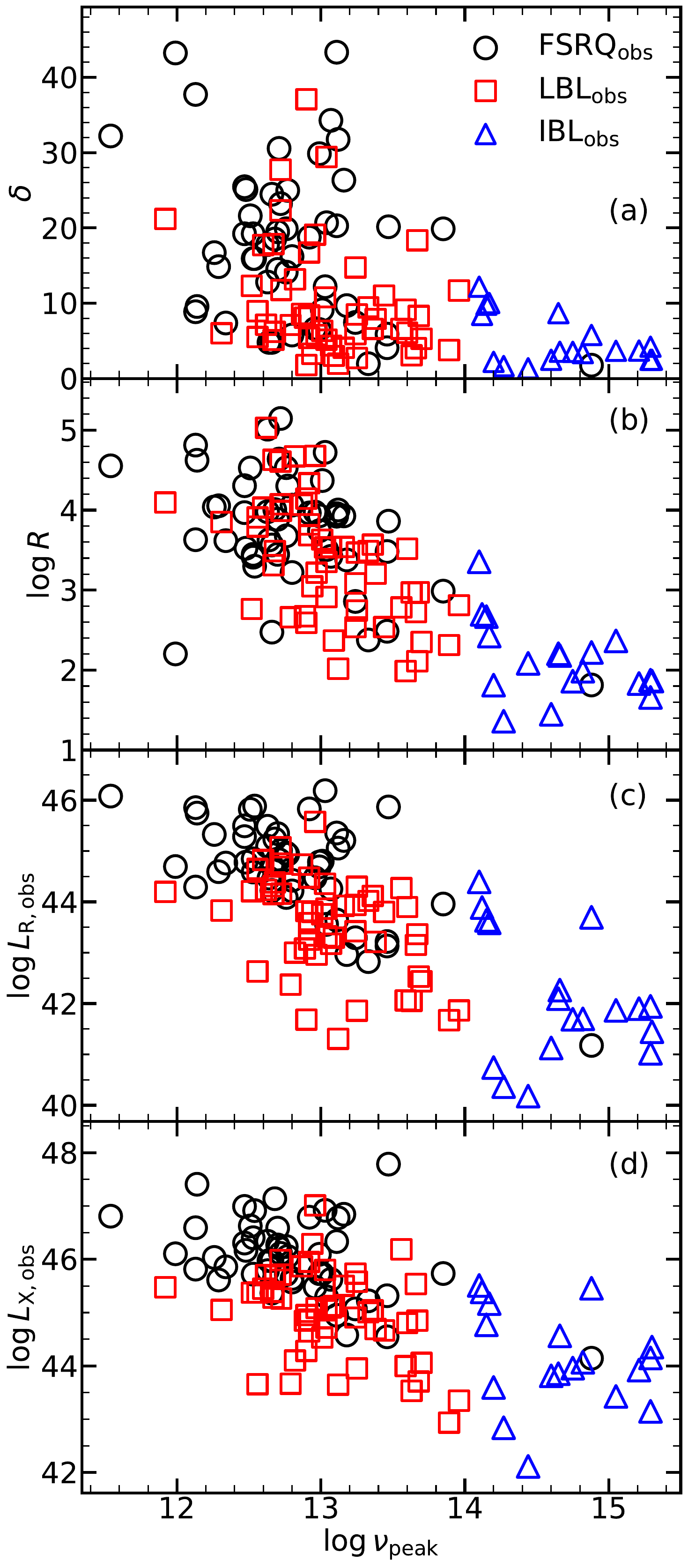}}
    \caption{The (a)-(d) show the synchrotron peak frequency ($\log \nu_{\rm{peak}}$) vs. the 5\,GHz Doppler factor ($\delta$), radio-loudness ($\log R$), observational 5\,GHz core radio luminosity ($\log L_{\rm{R,obs}}$), and 2--10\,keV X-ray luminosity ($\log L_{\rm{X,obs}}$), respectively. The black-circles, red-squares, and blue-triangles represent FSRQs, LBLs, and IBLs, respectively. The subscript "obs" or hollow-symbol denote the observational data.}
    \vspace{0.3cm}
    \label{logVpeak+other parameters}
\end{figure}
\subsection{The Doppler Boosting Effect for Blazars}\label{beaming}
\cite{Wu2007} and \cite{Nieppola_2008} found an anticorrelation between the Doppler factor $\delta$ and $\log \nu_{\rm{peak}}$ for the blazars with $\log \nu_{\rm{peak}}<15.3\,\rm{Hz}$, which was interpreted by \cite{Nieppola_2008} in the framework that the lower $\log \nu_{\rm{peak}}$ sources are being boosted more than higher $\log \nu_{\rm{peak}}$ sources. Their results imply that the lower $\log \nu_{\rm{peak}}$ blazars have more powerful jets than HBLs (i.e. $P_{\rm{jet}}\propto \delta \propto 1/\log \nu_{\rm{peak}}$). On the contrary, the HBLs show a horizontal line in $\delta-\log \nu_{\rm{peak}}$ relation \citep[see Figure\,13 in][]{Wu2007}, which may be due to the fact that they are weak jet sources. In addition, the anticorrelation between luminosities and $\log \nu_{\rm{peak}}$ \citep[e.g.,][]{Fossati1998unifying,Nieppola2006spectral,Nieppola_2008,Fan2016spectral,Yang2022spectral,Yang2022beaming} was also explained as the result of Doppler beaming effect.

Figure\,\ref{logVpeak+other parameters} (a)--(d) show the synchrotron peak frequency ($\log \nu_{\rm{peak}}$) versus the 5\,GHz Doppler factor ($\delta$), radio-loudness ($\log R$),  observational 5\,GHz core radio luminosity ($\log L_{\rm{R,obs}}$), and observational 2--10\,keV X-ray luminosity ($\log L_{\rm{X,obs}}$) for our sample, respectively. It is clear that all correlations above are negative. For Figure\,\ref{logVpeak+other parameters}\,(a), which is consistent with results of \cite{Wu2007} and \cite{Nieppola_2008} (see their Figure\,13 and Figure\,1, respectively). For Figure\,\ref{logVpeak+other parameters}\,(b), to the best of our knowledge, this is the first study of the relationship between $\log \nu_{\rm{peak}}$ and $\log R$. For Figure\,\ref{logVpeak+other parameters}\,(c-d), which are consistent with the results of \cite{Nieppola2006spectral}, \cite{Fan2016spectral}, and \cite{Yang2022spectral}. It is clear that the lower $\log\nu_{\rm{peak}}$ sources have a larger $\delta$, $\log R$, $L_{\rm{R,obs}}$ and $L_{\rm{X,obs}}$ (i.e., $1/\log \nu_{\rm{peak}} \propto \delta \propto \log R \propto L_{\rm{R,obs}} \propto L_{\rm{X,obs}}$). It is noting that the X-ray emissions have a contribution of jets (synchrotron radiation or inverse Compton process). Therefore, the physical explanation for those anticorrelation is the boosted consequence of Doppler beaming effect, where the lower $\log \nu_{\rm{peak}}$ sources have be more boosted, which causes a larger luminosity and radio-loudness \citep[also see][]{Yang2022spectral}.

From the above theoretical analysis, we have again enough confidence to believe that our sample is a typical strong jet sample with a strong Doppler boosting effect. Therefore, the influence of Doppler boosting effect must be considered in our following radio--X-ray correlation and FP analysis.

\subsection{Methods}\label{sec:FP}
It is clear that our sample has a strong beaming effect from the discussion and analysis in \S\,\ref{beaming}, the Doppler boosting effect therefore should be considered into the best fit of radio--X-ray correlation and the FP. In the beaming model, the observational flux density ($F_{\rm{obs}}$) is strongly boosted from  the intrinsic flux density ($F_{\rm{int}}$):
\begin{equation}
   F_{\nu,\rm{int}}=F_{\nu,\rm{obs}}\delta_{\nu}^{-q}
    \label{Eq2}
\end{equation}
where the $F_{\nu,\rm{int}}$, $F_{\nu,\rm{obs}}$ are the intrinsic flux density and the observational flux density at $\nu$ band, respectively. $q=2+\alpha$ for the continuous jets, where $\alpha$ is the spectral index. For the radio band, the typical value of $\alpha_{\rm{R}}=0$ \citep{Donato2001hard,Dong2023FP} is adopted in this work, and the X-ray spectral indices are calculated using the $\alpha_{\rm{X}}=\Gamma-1$ \citep{Lusso2010x}. Through Equation\,(\ref{Eq2}), we divided our data into 'observational data bin' and 'intrinsic data bin.'

For 5\,GHz radio Doppler factor ($\delta_{\rm{R}}$) of FSRQs and the 2--10\,keV X-ray Doppler factor ($\delta_{\rm{X}}$) of all blazars, which are extrapolated from the Doppler factors at other frequencies by using the assumption that Doppler factor depend on the emission frequency \citep{Fan1994bl}, the form of equation is as follow:
\begin{equation}
    \delta_{\nu}=\delta_{o}^{1+\frac{1}{8}\log\frac{\nu_o}{\nu}}
    \label{Eq3}
\end{equation}
where $\delta_o$ is the optical Doppler factor at the typical optical band of $\nu_o$ \citep[$\log\nu_o=14$\,Hz, see][]{Fan1994bl}. The 5\,GHz radio frequency and X-ray frequency at 2--10\,keV band are $\log\nu_{\rm{R}}=9.7$\,Hz and $\log\nu_{\rm{X}}=18.16$\,Hz, respectively. Hence, we can get an equation of $\delta_{\rm{X}}\approx \delta_{\rm{R}}^{1/3}$. Similarly, the 22\,GHz and 37\,GHz $\delta$ of FSRQs \citep{Hovatta2009doppler} are also converted to the 5\,GHz $\delta$.

To reexplore the FP for strong jet sources, a similar method in \cite{Merloni2003} is adopted in this work, which has the form $\log L_{\rm{R}}=\xi_{\rm{X}}\log L_{\rm{X}}+\xi_{\rm{M}}\log M_{\rm{BH}}+c_{0}$. In order to find the multivariate relation coefficients, we adopt the least $\chi^2$ approach in \cite{Merloni2003} and minimized the following statistic:
\begin{equation}
    \chi^{2}=\sum\limits_{i}\frac{(y_i-c_0-\xi_{\rm{X}}X_i-\xi_{\rm{M}}M_i)^2}{\sigma^2_{\rm{R}}+\xi_{\rm{X}}^2\sigma^2_{\rm{X}}+\xi_{\rm{M}}^2\sigma^2_{\rm{M}}}
    \label{Chi-square}
\end{equation}

where the $y_i=\log L_{\rm{5GHz}}$, $X_i=\log L_{\rm{2-10\,keV}}$, $M_i=\log M_{\rm{BH}}$ and $c_0$ is constant. For blazars, the uncertainties of the data mainly come from the nonsimultaneity between radio and X-ray emissions. Considering the systematic and the observed uncertainties, we adopt the typical isotropic uncertainties with $\sigma_{\rm{R}}=\sigma_{\rm{X}}=\sigma_{\rm{M}}=0.3$\,dex, following \cite{Merloni2003} and \cite{Xie2017FP}. Noticing, we used this approach to fit our intrinsic data bin and observational data bin, respectively. 
\begin{figure}
    \vspace{0.1cm}
    \centering{\includegraphics[scale=0.29]{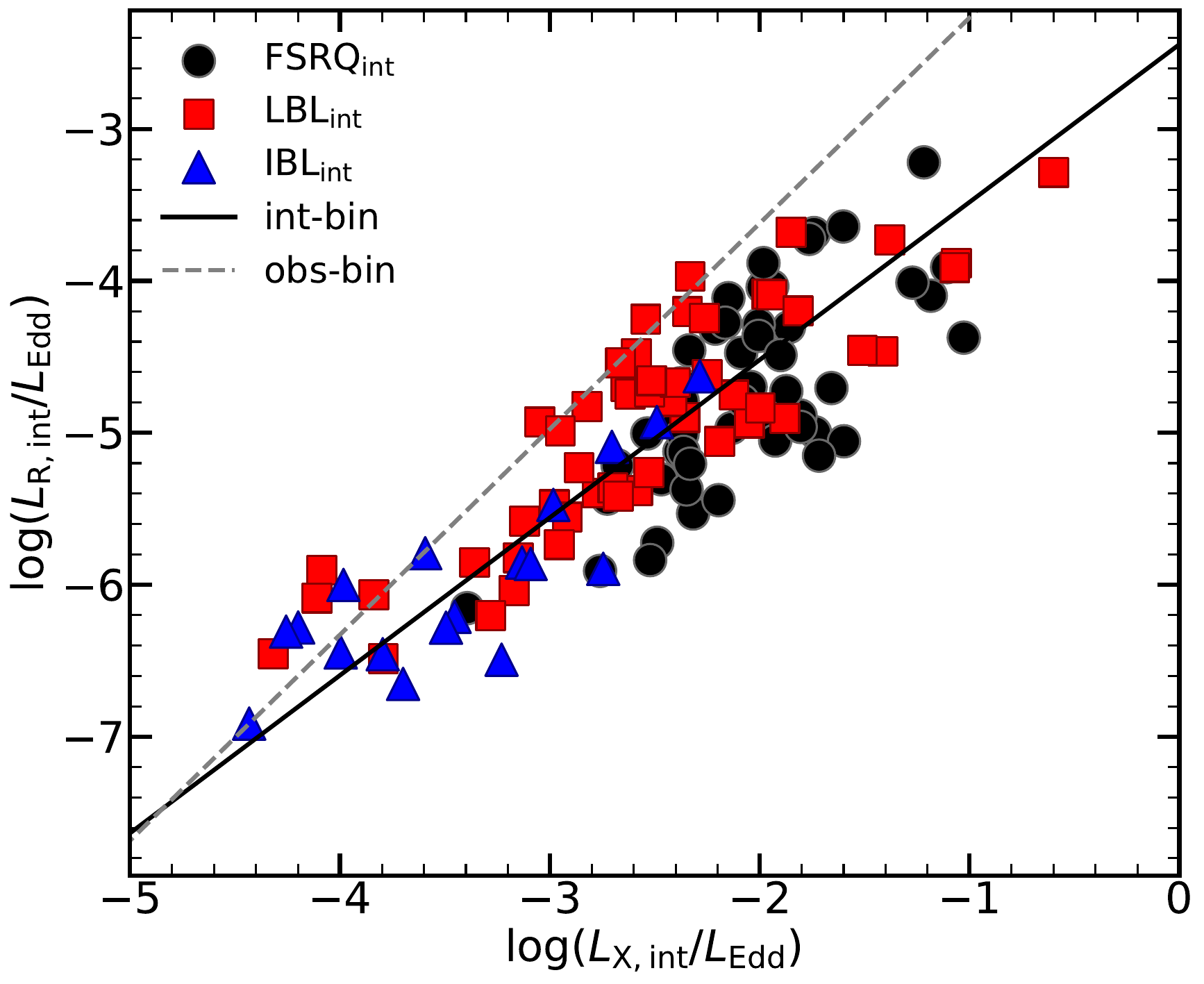}}
    \caption{The Eddington-scaled radio--X-ray correlation. Similarly, the black-circles, red-squares and blue-triangles represent FSRQs, LBLs and IBLs, respectively. The subscript 'int' or solid-symbol denote the intrinsic data. The black solid line represent best-fit line of the intrinsic data bin, the gray dashed line represent best-fit line of the observational data bin.}
    \vspace{0.1cm}
    \label{Lx/Ledd--Lr/Ledd}
\end{figure}
\subsection{Results}
By considering the Doppler beaming effect, we obtain the observational and intrinsic radio--X-ray correlation for our strong jet blazars. Figure\,\ref{Lx/Ledd--Lr/Ledd} shows our results, and their best fits are given as follows:
\begin{equation}
    \log(\frac{L_{\rm{R,obs}}}{L_{\rm{Edd}}})=(1.35\pm0.06)\log(\frac{L_{\rm{X,obs}}}{L_{\rm{Edd}}})-(0.91\pm0.10)
    \label{Eq5}
\end{equation}
with an intrinsic scatter of $\sigma_{\rm{int}}=0.42$\,dex, a Spearman correlation coefficient of $R$=0.89 and $P=2.02\times10^{-42}$.
\begin{equation}
    \log(\frac{L_{\rm{R,int}}}{L_{\rm{Edd}}})=(1.04\pm0.05)\log(\frac{L_{\rm{X,int}}}{L_{\rm{Edd}}})-(2.44\pm0.14)
    \label{Eq6}
\end{equation}
with an intrinsic scatter of $\sigma_{\rm{int}}=0.32$\,dex, a Spearman correlation coefficient of $R$=0.82 and $P=1.43\times10^{-29}$.

The intrinsic radio--X-ray correlation is roughly consistent with the theoretical prediction of the pure jet-dominated mode \citep[$L_{\rm{R}}\propto L_{\rm{X}}^{1.23}$, see][]{yuan2005radio}, while the Doppler beaming effect causes the observational radio--X-ray correlation to be steeper than the theoretical prediction above.

\begin{figure}
    \vspace{0.1cm}
    \centering{\includegraphics[scale=0.29]{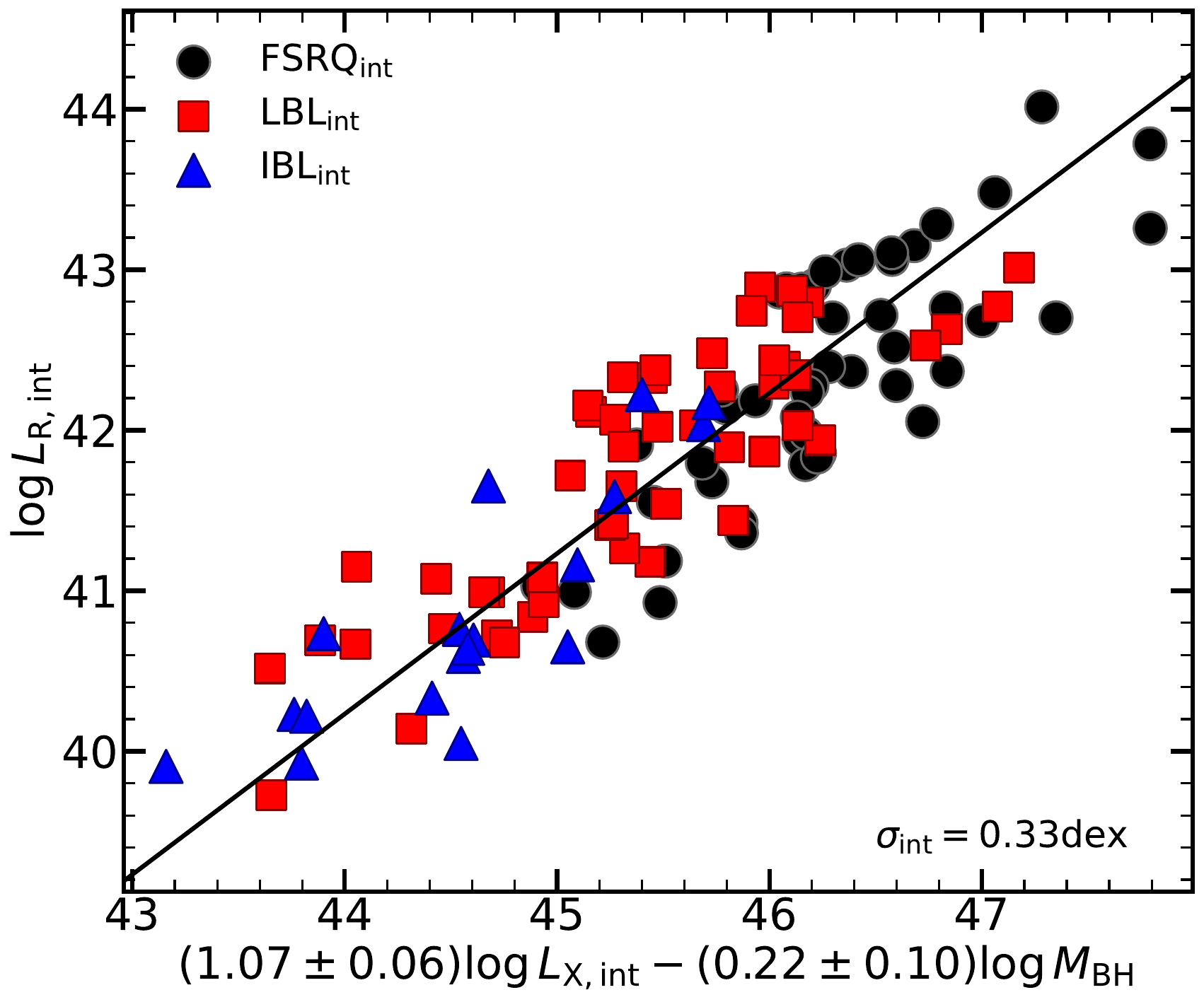}}
    \caption{The $\rm{FP_{int}}$ for our strong jet sources, where the notations are the same as that of Figure\,\ref{Lx/Ledd--Lr/Ledd}.}
    \vspace{0.1cm}
    \label{FP OF SAMPLES}
\end{figure}
Through Equation\,(\ref{Chi-square}), we obtain the observational FP ($\rm{FP_{obs}}$) and the intrinsic FP ($\rm{FP_{int}}$, see Figure\,\ref{FP OF SAMPLES}) for the strong jet sources. Their best fits are given as follows:
\begin{equation}
    \begin{split}
    \log L_{\rm{R,obs}}=&(1.39\pm0.06)\log L_{\rm{X,obs}}-(0.34\pm0.12)\log M_{\rm{BH}}\\
                        &-(15.98\pm2.18)
    \label{Eq7}
    \end{split}
\end{equation}
with a larger intrinsic scatter of $\sigma_{\rm{int}}=0.43$\,dex.
\begin{equation}
\begin{split}
   \log L_{\rm{R,int}}=&(1.07\pm0.06)\log L_{\rm{X,int}}-(0.22\pm0.10)\log M_{\rm{BH}}\\
                       &-(3.77\pm2.11)\label{Eq8} 
\end{split}
\end{equation}
with an intrinsic scatter of $\sigma_{\rm{int}}=0.33$\,dex.

We find that our $\rm{FP_{int}}$ is roughly consistent with the hybrid mode of jet+standard disk (SSD) in \cite{Bariuan2022FP} and \cite{Wang2024FP}, while the $\rm{FP_{obs}}$ is close to the FP of RL-AGNs of \cite{Li2008black} who attributed the steeper FP to the Doppler beaming effect.

\begin{figure}
    \vspace{0.3cm}
    \centering{\includegraphics[scale=0.31]{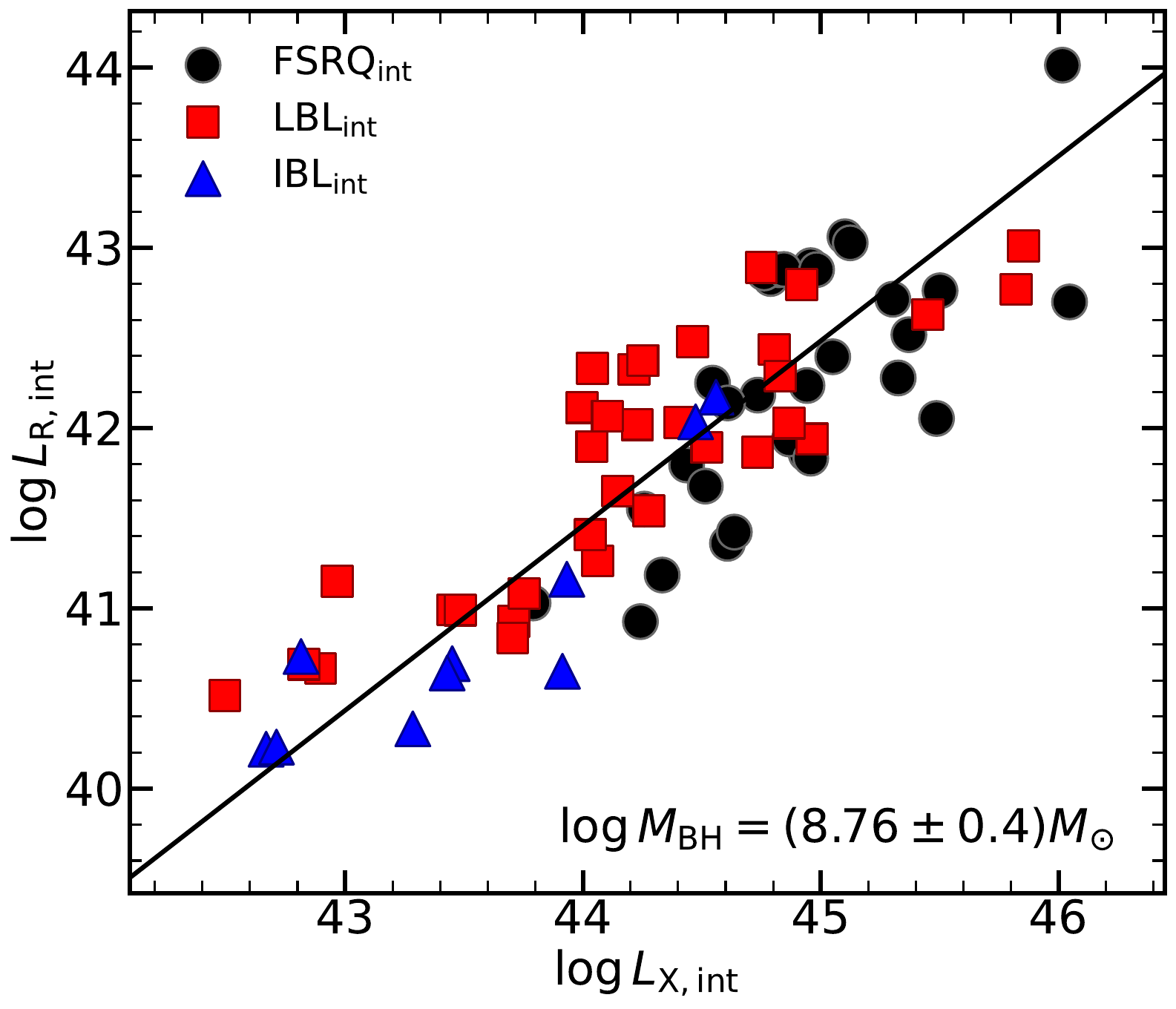}}
    \caption{The relation between 5 GHz radio and 2--10\,keV X-ray luminosity for strong jet subsample with BH mass $\log M_{\rm{BH}}=(8.80\pm0.4)M_{\odot}$. The notations are the same as that of Figure\,\ref{Lx/Ledd--Lr/Ledd}. The black solid line is the best fit of intrinsic data bin for Single-BH subsample.}
    \vspace{0.3cm}
    \label{BH-evolution}
\end{figure}

Following \cite{dong2015revisit}, in order to reinvestigate the radio--X-ray correlation and  eliminate the possible effect of mass, we further select a subsample with a narrow range of BH mass, which can simulate a single SMBH evolution for our strong jet sources. Finally, we selected 74 sources from table.\,\ref{table2}, which have the BH mass $\log M_{\rm{BH}}=\log \overline{M}_{\rm{BH}}\pm0.4=(8.76\pm0.4)M_{\odot}$ and the Eddington-ratio in a narrow range ($\lambda_{\rm{int}}=-4.5\sim-1$: which spans almost the entire range of our sample). The result is shown in Figure\,\ref{BH-evolution}, the best-fit linear regression between the intrinsic radio luminosity and the X-ray luminosity for our single-BH sample is the following:
\begin{equation}
    \log L_{\rm{R,int}}=(1.03\pm0.07)\log L_{\rm{X,int}}-(3.66\pm3.01)
    \label{Eq9}
\end{equation}
with an intrinsic scatter of $\sigma_{\rm{int}}=0.34$\,dex, a Spearman correlation coefficient of $R$=0.82 and $P=5.07\times10^{-19}$, which is roughly consistent with \cite{dong2015revisit} and \cite{Xie2017FP}.

\subsection{Correlation Tests}
The Malmquist biases are ubiquitous in astronomical surveys, and mitigating these biases is vital to perform astrophysical inference. Therefore, the luminosity--luminosity correlations should be tested for the possible presence of a spurious correlation introduced by their common dependence on the distance \citep{Merloni2003,Wang2006black,Li2008black}. Different from their method, we perform the Spearman partial correlation analysis to test the correlation between luminosities, as well as between the Eddington-luminosity-scaled luminosities. For the given three variables ($X$, $Y$, $Z$), the partial correlation coefficient between $X$ and $Y$, while keeping $Z$ as the third variable, can be expressed as
\begin{equation}
    R_{XY,Z}=\frac{R_{XY}-R_{XZ}R_{YZ}}{[(1-R_{XZ}^2)(1-R_{YZ}^2)]^{1/2}}
\end{equation}
where the $R_{XY}$ denotes the Spearman rank correlation coefficient between
 quantity $X$ and $Y$, and so on. Table.\,\ref{table1} shows the results of our test, namely, whether the correlation between $X$ and $Y$ is intrinsic or only introduced by a third variable $Z$. The null hypothesis will be rejected when its probability less than the significance level (i.e., $0.05$).

\begin{figure*}
    \centering{\includegraphics[scale=0.31]{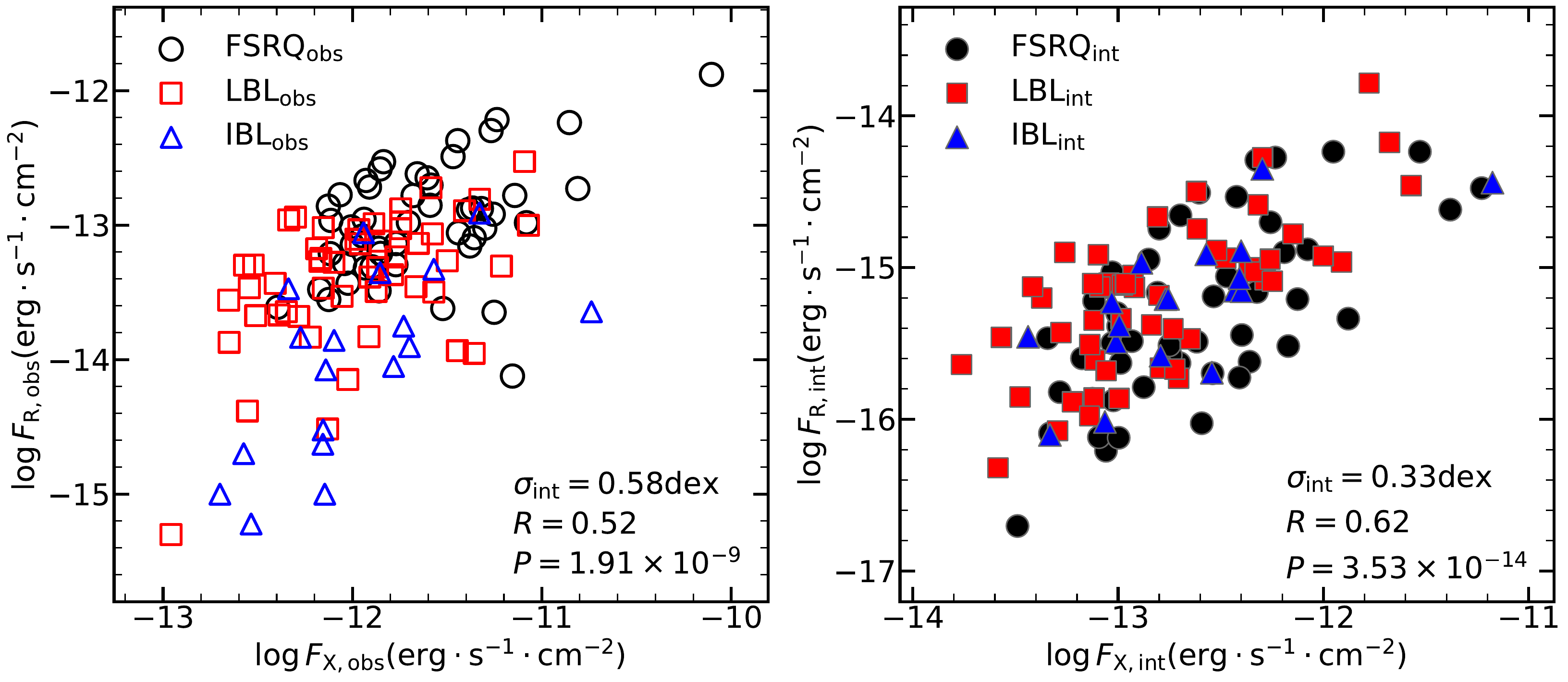}}
    \caption{The left panel and right panel show the observational 5\,GHz core radio flux densities vs. the observational 2--10\,keV X-ray flux densities and the intrinsic 5\,GHz core radio flux densities vs. the intrinsic 2--10\,keV X-ray flux densities, respectively. The notations are the same as that of Figure\,\ref{logVpeak+other parameters} and Figure\,\ref{Lx/Ledd--Lr/Ledd}.}
    \label{Fr-Fx}
\end{figure*}

For our correlations between luminosities, the same dependence on the distance always confuses the intrinsic physical relation. This is the main reason why \cite{Bregman2005absence} thinks the FP is a distance artifact. The partial correlation analysis indeed proves that luminosities are still strongly correlated, even if the effect of distance is included. However, the significance level of radio--X-ray correlations becomes weaker when the distance is taken into account compared to BH mass. As in \cite{Wang2006black}, in order to avoid the distance effect, we test the existence of the intrinsic correlation between radio and X-ray emissions by comparing the radio and X-ray flux density. In Figure\,\ref{Fr-Fx}, we plot the 5\,GHz radio flux density versus the 2--10\,keV X-ray flux density. The observational and intrinsic radio--X-ray emissions are as follows:
\begin{equation}
    \log F_{\rm{R,obs}}=(1.60\pm0.16)\log F_{\rm{X,obs}}+(5.58\pm1.95)
\end{equation}
with an intrinsic scatter of $\sigma_{\rm{int}}=0.58$\,dex, a Spearman correlation coefficient of $R$=0.52 and $P=1.91\times10^{-9}$.
\begin{equation}
    \log F_{\rm{R,int}}=(1.06\pm0.09)\log F_{\rm{X,int}}-(1.82\pm1.13)
\end{equation}
with an intrinsic scatter of $\sigma_{\rm{int}}=0.33$\,dex, a Spearman correlation coefficient of $R$=0.62 and $P=3.53\times10^{-14}$.

It is clear that the correlation between radio and X-ray emissions of our sample really exists, even if the significance level has declined compared to the correlation between luminosities. However, the intrinsic scatter ($\sigma_{\rm{int}}$) becomes smaller, and the significance level of radio--X--ray correlation becomes stronger when the Doppler factor is taken into account.

It can be found from the right panel of Figure\,\ref{Fr-Fx} that the relation of $\log F_{\rm{R,int}}-\log F_{\rm{X,int}}$ of FSRQs, LBLs, and IBLs show three different intercepts. Considering jet-dominated X-ray and radio emissions, the radio and X-ray flux follows the scaling relation, which is obtained from \cite{Heinz2004constraints},
\begin{equation}
    F_{\rm{R}}\propto M_{\rm{BH}}^{\frac{2p+13-(2+p)\alpha_{\rm{\scalebox{0.37}{R}}}}{2p+8}}\dot{m}^{\frac{2p+13+(p+6)\alpha_{\rm{\scalebox{0.37}{R}}}}{2(p+4)}}
    \label{Eq13}
\end{equation}
\begin{equation}
    F_{\rm{X}}\propto M_{\rm{BH}}^{(2-\alpha_{\rm{x}})/2}\dot{m}^{(5-3\alpha_{\rm{x}})/2}
    \label{Eq14}
\end{equation}
where the $p$ is the electron spectral index produced by the acceleration process \citep[see][]{Merloni2003,Heinz2004constraints}. Combining Equation\,(\ref{Eq13}) with Equation\,(\ref{Eq14}), we can obtain the desired relation:
\begin{equation}
    F_{\rm{R}}\propto M_{\rm{BH}}^{\frac{(2p+13-(2+p)\alpha_{\rm{\scalebox{0.37}{R}}})(p-1-\alpha_{\rm{x}})-2\alpha_{\rm{\scalebox{0.37}{R}}}}{(p+4)(2p+1-3\alpha_{\rm{x}})}}F_{\rm{X}}^{\frac{2p+13+(p+6)\alpha_{\rm{\scalebox{0.37}{R}}}}{(p+4)(2p+1-3\alpha_{\rm{x}})}}
\end{equation}
We can further get the relation of $\log F_{\rm{R}}$ and $\log F_{\rm{R}}$, that is,
\begin{equation}
    \begin{split}
    &\log F_{\rm{R}}\propto \frac{2p+13+(p+6)\alpha_{\rm{R}}}{(p+4)(2p+1-3\alpha_{\rm{X}})}\log F_{\rm{X}}+ \\
    &\frac{(2p+13-(2+p)\alpha_{\rm{R}})(p-1-\alpha_{\rm{X}})-2\alpha_{\rm{R}}}{(p+4)(2p+1-3\alpha_{\rm{X}})}\log M_{\rm{BH}}
    \label{Eq16}
    \end{split}
\end{equation}
From Equation\,(\ref{Eq16}), the difference of intercepts for the three different classes of objects may be related to the BH mass, the electron spectral index, the radio spectral index, and the X-ray spectral index.

\begin{deluxetable*}{c c c c c c}
\tablecaption{The Spearman Partial Correlation Analysis\label{table1}}
		\renewcommand{\tabcolsep}{19pt}
    \tablehead{$X$ & $Y$ & $Z$ & \colhead{Number} & \colhead{$R_{XY,Z}$} & \colhead{$P_{\rm{null}}$} }

	\startdata
	$\log L_{\rm{X,int}}$ & $\log L_{\rm{R,int}}$ & $\log D$&119& 0.58 & $6.99\times10^{-12}$ \\
        $\log (L_{\rm{X,int}}/L_{\rm{Edd}})$ & $\log (L_{\rm{R,int}}/L_{\rm{Edd}})$ & $\log D$&119& 0.73 & $4.28\times10^{-21}$ \\
        $\log L_{\rm{X,int}}$ & $\log L_{\rm{R,int}}$ &None&119& 0.89 & $3.02\times10^{-42}$\\
	$\log L_{\rm{X,int}}$ & $\log L_{\rm{R,int}}$ &$\log M_{\rm{BH}}$&119& 0.86 & $7.96\times10^{-35}$\\
        $\log (L_{\rm{X,int}}/L_{\rm{Edd}})$ & $\log (L_{\rm{R,int}}/L_{\rm{Edd}})$ & $\log M_{\rm{BH}}$&119& 0.86 & $7.96\times10^{-35}$ \\
        \enddata
    
    \tablecomments{Col\,(1): Variable $X$. Col\,(2): Variable $Y$. Col\,(3): Variable $Z$. Col\,(4): Number of the sources. Col\,(5): The partial correlation coefficient $R_{XY,Z}$. Col\,(6): The probability of null hypothesis $P_{\rm{null}}$.}
    
\end{deluxetable*}

\section{Discussion}\label{Discussion}
\subsection{The Influence of Doppler Beaming Effect on Radio--X-Ray Correlation and FP}
In some of the previous work studying FP of RL-AGNs\citep[e.g.,][]{Wang2006black,Li2008black,Bariuan2022FP}, their samples were doped with a few blazars, which may cause a spurious result. In this paper, we will emphasize the importance of considering the Doppler beaming effect for the strong jet blazars. It is well known that blazars have a nonnegligible Doppler beaming effect, especially for blazars with $\log \nu_{\rm{peak}}<15.3$, as confirmed in \S\,\ref{beaming}. From Figure\,\ref{Lx/Ledd--Lr/Ledd}, Equation\,(\ref{Eq7}) and (\ref{Eq8}), we can find that the radio--X-ray correlation and FP become shallower by considering the Doppler beaming effect, which is consistent with \cite{Zhang2024FP}. The plausible explanations are given as the following. On average, $\delta_{\rm{FSRQs}}>\delta_{\rm{LBLs}}>\delta_{\rm{IBLs}}$ and $L_{\rm{FSRQs}}>L_{\rm{LBLs}}>L_{\rm{IBLs}}$ (also see Figure\,\ref{logVpeak+other parameters}, $\delta \propto L_{\rm{R}} \propto L_{\rm{X}}$), it is clear that a larger $\delta$ causes a larger boosting in luminosities \citep[also see][]{Nieppola_2008,Yang2022spectral}. Therefore, we can see that the gap between the black solid line and the gray dotted line in Figure\,\ref{Lx/Ledd--Lr/Ledd} is getting bigger. Moreover, we also find the scatter of radio--X-ray correlation and FP get smaller by taking the Doppler factor into account. However, the significant level of observational radio--X-ray correlation is larger than that of intrinsic radio--X-ray correlation, which is due to the Doppler beaming effect and the distance effect stretching the range of luminosities. The significant level of intrinsic correlation between radio and X-ray emissions increases when eliminating the effect of distance (see Figure\,\ref{Fr-Fx}).

There is a noteworthy issue here, the Doppler factor is an unobserved quantity, and our Doppler factors are estimated by different method in the different literatures \citep[see][]{Hovatta2009doppler,Wu2014some,Ye2021unification}. However, we predict this does not have a significant effect on our results, owing to the anticorrelation between the Doppler factor and $\log \nu_{\rm{peak}}$ intrinsically exist even if the Doppler factors are calculated in same method \citep{Wu2007,Nieppola_2008,Yang2022beaming}. Therefore, our results are consistent with the theoretical expectations rather than the results being accidental.

\subsection{A Significant Impact of Eddington luminosity ratios on Radio--X-Ray Correlation and FP for the Strong Jet Sources?}
Previous works have suggested that the radio--X-ray correlation and FP depend on $\lambda_{\rm{Edd}}$ \citep[see][]{yuan2005radio,yuan2009revisiting,dong2014new,Xie2017FP,Wang2024FP}, which was interpreted as the transition of accretion mode (e.g., $\lambda_{\rm{Edd}}<-6$ for jet-dominated mode, $-6<\lambda_{\rm{Edd}}<-3$ for ADAF mode, $\lambda_{\rm{Edd}}>-3$ for disk mode). However, their works mainly focus on the LLAGNs instead of brighter blazars. Therefore, we want to examine whether the radio--X-ray correlation and FP of our blazars have a significant dependence on $\lambda_{\rm{Edd}}$. However, our blazar sample has a relatively narrow range in the Eddington-ratio (see Figure\,\ref{fig:histogramsources}\,(i)), which will not allow us to divide our sample into several subsample with a narrower range of $\lambda_{\rm{Edd}}$. Because, for the linear regression statistics, within the narrower range of $\lambda_{\rm{Edd}}$, the radio-X-ray correlations and FP will produce erroneous slopes; we suggest that the range of $\lambda_{\rm{Edd}}$ should be larger than $3$. Before performing the discussion, we need to emphasize that previous works have demonstrated that our blazars ($\log \nu_{\rm{peak}}<15.3$) are strong jet sources that exhibit the radiatively efficient accretion \citep{Meyer2011blazar,Keenan2021relativistic}. Therefore, they should favor the jet-dominated mode within the theoretical expectation. As expected, our sources tightly traced the jet-dominated mode and do not show a significant bias along the whole track (see Figure\,\ref{Lx/Ledd--Lr/Ledd}, \ref{FP OF SAMPLES}, \ref{BH-evolution} and \ref{Fr-Fx}). Statistically, by considering the Doppler beaming effect, we find that our intrinsic correlations have a small scatter ($\sigma_{\rm{int}}\le 0.34$) and a relative good significant level. Those evidences above imply the radio--X-ray correlation and FP of our strong jet sources do not have a significant dependence on $\lambda_{\rm{Edd}}$.

\begin{figure}
    \vspace{0.3cm}
    \centering{\includegraphics[scale=0.243]{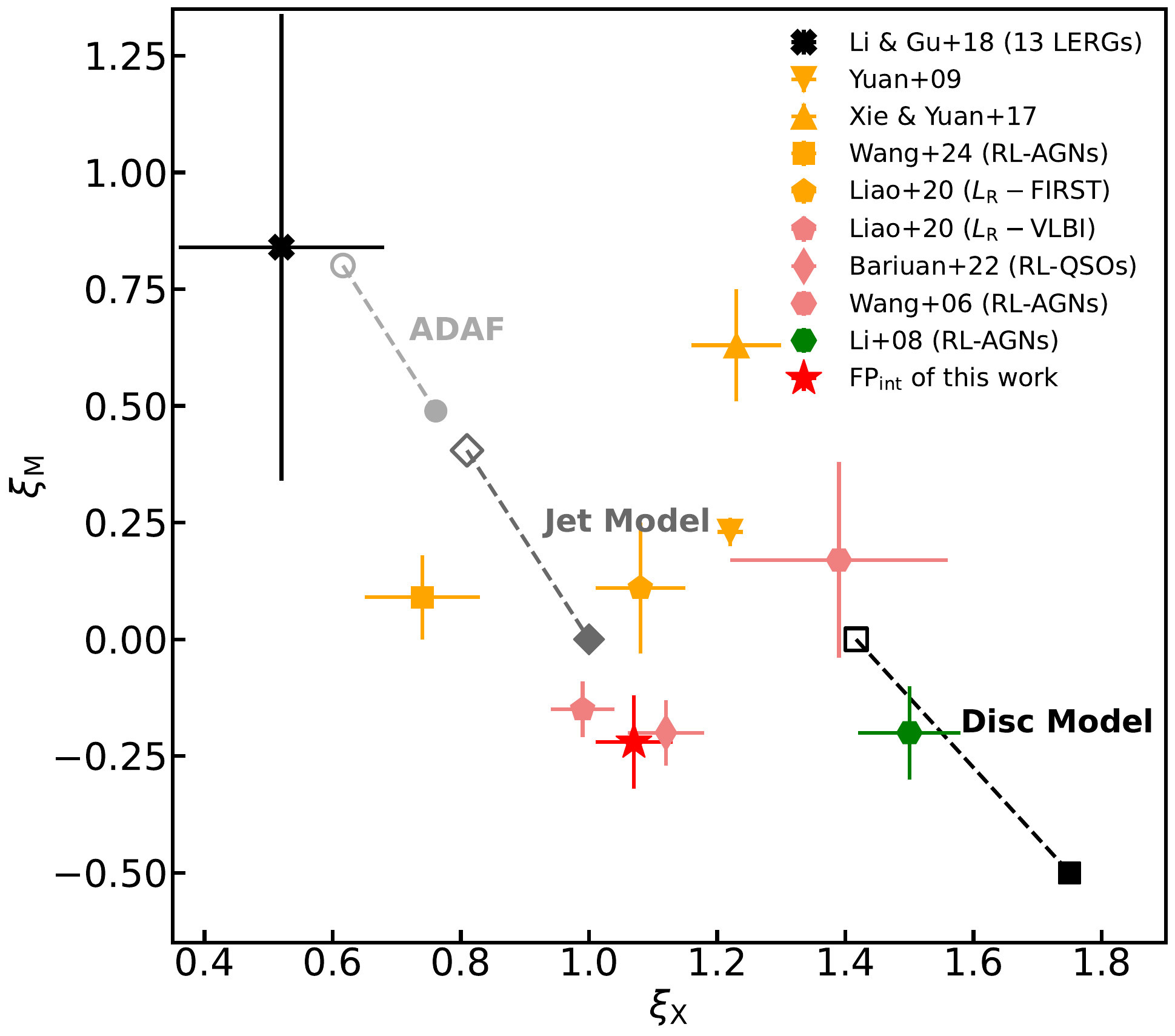}}
    \caption{A comparison the best-fit correlation coefficients $\xi_{\rm{X}}$ and $\xi_{\rm{M}}$ of FP for radio-loud samples, include several previous works and this work. The black-cross is FP of RL-AGNs (13 LERGs) in \cite{li2018black}. The orange inverted-triangle, orange triangle, orange square, and orange pentagon are FP of \cite{yuan2009revisiting}, FP of \cite{Xie2017FP}, $\rm{FP_{RL-AGNs}}$ of \cite{Wang2024FP}, and $\rm{FP_{FIRST}}$ of \cite{Liao2020x}, respectively. Note where the $L_{\rm{R}}-\rm{VLBI}$ and $L_{\rm{R}}-\rm{FIRST}$ represent the core radio emissions taken from VLBI and FIRST observation, respectively. The light-red pentagon, light-red diamond, and light-red hexagon are $\rm{FP_{VLBI}}$ of \cite{Liao2020x}, $\rm{FP_{RL-QSOs}}$ of \cite{Bariuan2022FP}, and $\rm{FP_{RL-AGNs}}$ of \cite{Wang2006black}, respectively. The green hexagon is $\rm{FP_{RL-AGNs}}$ of \cite{Li2008black}. The red star is intrinsic FP ($\rm{FP_{int}}$) in this work. In addition, the gray-circles, dark-gray diamonds, and black squares represent the theoretically predicted correlation coefficients of ADAF, jet, and disk mode from \cite{Merloni2003}, respectively. Empty symbols are results for a radio spectral index $\alpha_{\rm{R}}=0$ and filled symbols for $\alpha_{\rm{R}}=0.5$, and three dashed lines represent different possible correlation coefficients due to variations of radio emission.}
    \vspace{0.3cm}
    \label{Space}
\end{figure}
\subsection{The FP and X-Ray Emissions' Origin of Strong Jet Sources}
Theoretically, the collimated relativistic jets are produced from the innermost regions of the accretion disk, meaning that jet variables depend on the accretion rate ($\dot{M}$), and the scaled correlation was given by \cite{Heinz2003non}
\begin{equation}
    L_{\rm{R}}\propto{\dot{M}^{1.4}}
\end{equation}
In addition, the X-ray emissions are widely believed come from the inverse Compton scatter of soft photons and the form of the X-ray depends on $\dot{M}$ as the following: 
\begin{equation}
    L_{\rm{X}}\propto{\dot{M}^{q}}
\end{equation}
Therefore, the correlation of radio--X-ray can be written as 
\begin{equation}
     L_{\rm{R}}\propto{L_{\rm{X}}^{1.4/q}} 
\end{equation}
where $q$ value is corresponding to different mode; $q=2.3$, $q=1$ and $q=1.1$ can be adopted for the radiative inefficiency flow \citep{Merloni2003}, the radiative efficiency flow \citep{dong2014new} and a jet-dominated mode \citep{yuan2005radio}, respectively.

Our intrinsic radio--X-ray correlation ($L_{\rm{R,int}}\propto L_{\rm{X,int}}^{1.04}$) favors the jet-dominated mode in \cite{yuan2005radio}, which implies that the X-ray emissions of strong jet sources mainly come from jets.  For observational radio--X-ray correlation ($L_{\rm{R,obs}}\propto L_{\rm{X,obs}}^{1.35}$: the black dashed line in Figure\,\ref{Lx/Ledd--Lr/Ledd}), which is steeper than the pure jet-dominated mode ($L_{\rm{R}}\propto L_{\rm{X}}^{1.23}$), the physical reason may be that there is an extra contribution from the Doppler beaming effect. In addition, our result (see Figure\,\ref{BH-evolution}) simulating the evolution of a single-BH for the strong jet sources still follows the jet-dominated mode \citep[see][]{dong2015revisit,Xie2017FP}, which further strengthens the conclusions of the present work.

The measured correlation coefficients ($\xi_{\rm{X}}$, $\xi_{\rm{M}}$) from FP can provide insight into the different emission mechanisms within the BHs, where the ratio of these observables corresponds to the predicted values from different emission modes \citep{Merloni2003,Plotkin12}. In order to reexplore and refine the FP and X-ray emissions' origin of RL-AGNs, we plotted the $\xi_{\rm{X}}-\xi_{\rm{M}}$ diagram of FP of previous radio-loud works and this work in the parameter space of best-fit correlation coefficient (see Figure\,\ref{Space}). We now classify accretion physics for FP of all radio-loud samples according to the $\xi_{\rm{X}}-\xi_{\rm{M}}$ cartoon diagram in \cite{Wang2024FP}, detailedly see their Figure\,6. Combining our Figure\,\ref{Space} with their Figure\,6, we can find that the FP of \cite{li2018black} are consistent with ADAF mode in \cite{Merloni2003}. The FP of \cite{Wang2024FP}, \cite{Xie2017FP}, \cite{yuan2009revisiting} and $\rm{FP_{FIRST}}$ in \cite{Liao2020x} agree with the pure jet-dominated mode in \cite{yuan2005radio}, while the FP of \cite{Wang2006black}, \cite{Bariuan2022FP}, $\rm{FP_{VLBI}}$ in \cite{Liao2020x} and our $\rm{FP_{int}}$ favor the hybrid mode of jet+SSD (but the contribution of jets is probably bigger). The FP of \cite{Li2008black} are located in the region of Disk mode, \cite{Li2008black} thought the X-ray emissions of RL-AGNs come from accretion disk (support disk mode), and the reason for steeper FP is due to the Doppler beaming effect. Here, our $\rm{FP_{obs}}$ are close to their FP, but it is clear that the Doppler beaming effect causes $\rm{FP_{obs}}$ to bias the disk mode. Because our strong jet blazars have a $\log \nu_{peak}<15.3$, according to the SED \citep{Fossati1998unifying,Donato2001hard}, their X-ray emissions should come from the synchrotron emission tail or inverse Compton emissions in jet \citep[also see][]{Nieppola2006spectral,Wu2007,Wu2014some,Fan2016spectral,Yang2022spectral}. Therefore, it is difficult for us to circumvent the X-ray emissions come from the jets. To sum up, the most plausible explanation for our $\rm{FP_{obs}}$ is simply the joint result of jet-dominated flow and the Doppler beaming effect. 

For the FP of \cite{li2018black}, their LERGs samples are RL-AGNs but still consistent with ADAF mode instead of jet-dominated mode. The best physical explanation is that LERGs are weak jet sources and dominated by radiatively inefficient accretion flows \citep{Hardcastle2007hot,Meyer2011blazar,Keenan2021relativistic}. Therefore, the radio-loud source is not necessarily a strong jet source. So care must be taken when placing RL-AGNs on FP \citep[also see][]{Hardcastle2009active}. 

\subsection{The Dependence of FP on the BH Mass}
From Figure\,\ref{Lx/Ledd--Lr/Ledd}, it can be found that the slope of the radio–X-ray correlation $\xi_{\rm{X}}$ is 1.04. Considering the BH mass, we can obtain the FP, where the $\xi_{\rm{X}}\sim 1.07$ (see Figure\,\ref{FP OF SAMPLES}). The above results suggest that the BH mass has a weak influence on the FP. Additionally, the $\xi_{\rm{M}}\sim-0.22$ of $\rm{FP_{int}}$, which is similar to that of \cite{Wang2006black}, \cite{yuan2009revisiting}, \cite{Liao2020x} and \cite{Bariuan2022FP}, also imply the BH mass has a weak influence on the FP and roughly agree with the theoretical prediction of jet-dominated mode ($0\leq \xi_{\rm{M}}\leq 0.41$: see Figure\,\ref{Space}). To test this, we perform the partial correlation analysis by taking the BH mass into account (see table.\,\ref{table1}). From table.\,\ref{table1}, we can find that the correlation coefficient ($R_{XY,Z}$) between $\log L_{\rm{R,int}}$ and $\log L_{\rm{X,int}}$ only decreases from $0.89$ to $0.86$, which implies the our FP is not sensitive to the BH mass. Furthermore, we analyze the BH mass dependence of ADAF and SSD mode; \cite{Merloni2003} and \cite{li2018black} found the radiatively inefficient FP with the $\xi_{\rm{M}}\sim0.78$ and $\xi_{\rm{M}}\sim0.84$, respectively, which roughly agree with the theoretical prediction of ADAF mode ($0.49\leq \xi_{\rm{M}}\leq 0.80$) in Figure\,\ref{Space}, suggesting that the ADAF mode has a strong dependence on BH mass. \cite{dong2014new} and \cite{Li2008black} found the radiatively efficient FP with the $\xi_{\rm{M}}\sim-0.22$ and $\xi_{\rm{M}}\sim-0.20$, respectively, which are consistent with the theoretical prediction of SSD mode ($-0.50\leq \xi_{\rm{M}}\leq 0$) in Figure\,\ref{Space}, suggesting that the SSD mode has a weak dependence on BH mass. Therefore, a possibly reasonable interpretation is that the BH mass dependence of FP is regulated by the accretion mode.

\section{Summary}\label{lastpage}
It is  widely believed that the X-ray emissions of strong jet sources (e.g., RL-AGNs and Blazars) mainly from the nonthermal emission in relativistic jets, but it is still controversial. Currently, there are two opposite views for this issue; a popular view \citep{Kording06RefiningA&A,De-Gasperin2011testing,Plotkin12,Liao2020x,dong_2021,Dong2023FP,Bariuan2022FP,Wang2024FP} supports the jet-dominated mode in \cite{yuan2005radio}; ones are opposite \citep{li2018black,li2021origin} that favor the ADAF mode. Motivated by these issues, we constructed a strong jet blazar sample including  $50$ FSRQs, $51$ LBLs and $18$ IBLs to study the radio--X-ray correlation and the FP. Previous studies have suggested the radio--X-ray correlation and FP depend on $\lambda_{\rm{Edd}}$ but they mainly focus on general sources (e.g. LLAGNs); such studies are absent in bright blazars. Moreover, we take the Doppler factor into account due to the fact that our strong jet blazars have a nonnegligible Doppler beaming effect. Our main results can be summarized as the following: 

(1) By considering the Doppler beaming effect, we find the intrinsic radio--X-ray correlation of strong jet sources is $L_{\rm{R,int}}\propto L_{\rm{X,int}}^{1.04}$,  which favor the jet-dominated mode. This correlation remains even after removing the effect of distance, suggesting that the X-ray emissions of strong jet sources mainly come from jets. In addition, the simulation of the evolution of a single strong jet BH is still following the jet-dominated trace, which further strengthens our conclusions. The observational radio--X-ray correlation ($L_{\rm{R,obs}}\propto L_{\rm{X,obs}}^{1.35}$) suggests that the steeper slope may be due to the extra contribution from the Doppler beaming effect.

(2) Similarly, we obtain the $\rm{FP_{int}}$ by considering the Doppler beaming effect: $\log L_{\rm{R,int}}=(1.07\pm0.06)\log L_{\rm{X,int}}-(0.22\pm0.10)\log M_{\rm{BH}}-(3.77\pm2.11)$, which is interpreted by the hybrid mode of jet+SSD, implying that the X-ray emissions of strong jet sources are dominated by jets, but there may also be a small contribution from the disk.

(3) Our sources tightly traced the jet-dominated mode and do not show a significant bias along the whole track. In statistics, the intrinsic correlations have a small scatter ($\sigma_{\rm{int}}\le 0.34$) and a relatively good significant level. Those evidences above imply the radio--X-ray correlation and FP of our strong jet sources have not a significant dependence on $\lambda_{\rm{Edd}}$. The plausible reason is that our blazars are the strong jet sources; thus, they should favor the jet-dominated mode in the theoretical analysis.

\section*{Acknowledgements}
This work is supported by the NSFC (12363005, 2022SKA0130104), the Foundation of Guizhou Provincial Education Department ((2020)003, QJHKYZ[2021]296), the Scientific Research Project of the Guizhou Provincial Education (KY[2022]132, KY[2022]137), Major Science and Technology Program of Xinjiang Uygur Autonomous Region (2022A03013-4) and the Science and the Technology Foundation of Guizhou Province (Nos. ZK[2022]304, QKHJC[2020]1Y018).

\bibliography{Paper}{}
\bibliographystyle{aasjournal}

\begin{longrotatetable}
\begin{deluxetable*}{cccccccccccccc}
		\tablecaption{The properties of Blazars\label{table2}}
		\renewcommand{\tabcolsep}{6.5pt}
		\tablehead{
			\colhead{IAU Name} & \colhead{$z$} & 
			\colhead{$\log \nu_{\rm{peak}}$} & \colhead{Type} & 
			\colhead{$\delta$} & \colhead{$\Gamma$} &  \colhead{$\log L_{\rm{X,obs}}$} & 
			\colhead{$\log L_{\rm{X,int}}$} & \colhead{$\log L_{\rm{R,obs}}$} & \colhead{$\log L_{\rm{R,int}}$} & \colhead{$\log M_{\rm{HB,dyn}}$} & \colhead{Refs.} & \colhead{$\log R$} & \colhead{$\log \lambda_{\rm{int}}$} \\ 
			\colhead{(1)} & \colhead{(2)} & \colhead{(3)} & \colhead{(4)} & 
			\colhead{(5)} & \colhead{(6)} &
			\colhead{(7)} & \colhead{(8)} & \colhead{(9)} & \colhead{(10)} & \colhead{(11)} & \colhead{(12)} & \colhead{(13)} & \colhead{(14)}
		} 
		\startdata
0007$+$106&0.089&14.88&ISF&1.77&1.75&44.143&43.916&41.176&40.680&8.29&1, 2&1.81&$-2.488$\\
0106$+$013&2.099&12.51&LSF&21.61&1.45&46.626&45.536&45.819&43.150&9.51&3, 4&4.53&$-2.088$\\
0133$+$476&0.859&12.66&LSF&24.47&1.40&45.979&44.868&44.717&41.940&8.73&5, 6&3.56&$-1.976$\\
0149$+$218&1.320&12.80&LSF&5.77&1.69&45.640&44.958&44.424&42.902&8.83&3, 7&4.06&$-1.986$\\
0212$+$735&2.367&12.14&LSF&9.57&1.51&47.407&46.587&45.745&43.784&9.58&8, 9&4.62&$-1.107$\\
0234$+$285&1.213&12.76&LSF&19.89&1.40&46.239&45.200&44.963&42.366&9.22&3, 10&4.53&$-2.134$\\
0306$+$102&0.863&12.76&LSF&14.13&1.65&45.715&44.699&44.086&41.786&7.77&3, 9&3.68&$-1.185$\\
0333$+$321&1.263&13.16&LSF&26.34&1.57&46.845&45.628&45.208&42.367&9.25&11, 12&3.93&$-1.736$\\
0355$+$508&1.510&12.54&LSF&15.93&1.47&46.913&45.923&45.884&43.480&9.67&8, 13&3.30&$-1.860$\\
0420$-$014&0.915&12.77&LSF&24.98&2.10&46.053&44.608&44.935&42.140&9.03&14, 6&4.30&$-2.536$\\
0440$-$003&0.844&12.29&LSF&14.86&1.73&45.612&44.545&44.594&42.250&8.81&3, 15&4.05&$-2.379$\\
0458$-$020&2.286&12.70&LSF&19.49&1.52&46.585&45.502&45.342&42.762&8.66&16, 9&3.89&$-1.272$\\
0528$+$134&2.070&12.13&LSF&37.72&1.58&46.593&45.237&45.853&42.700&9.80&17, 18&4.81&$-2.677$\\
0552$+$398&2.363&11.54&LSF&32.21&1.63&46.808&45.486&46.080&43.064&9.74&8, 19&4.55&$-2.368$\\
0605$-$085&0.872&12.13&LSF&8.87&1.42&45.815&45.050&44.291&42.396&8.98&3, 9&3.63&$-2.043$\\
0642$+$449&3.396&13.03&LSF&12.20&1.49&46.918&46.016&46.186&44.014&9.12&3,  &4.72&$-1.218$\\
0736$+$017&0.191&13.18&LSF&9.68&  &44.579&43.731&42.963&40.990&8.00&20, 6&3.38&$-2.383$\\
0804$+$499&1.436&11.99&LSF&43.23&1.69&46.103&44.637&44.695&41.423&8.84&3, 21&2.20&$-2.317$\\
0827$+$243&0.941&12.53&LSF&15.93&1.46&45.721&44.735&44.587&42.183&9.01&22, 21&3.41&$-2.389$\\
0836$+$710&2.218&13.47&LSF&20.15&1.46&47.785&46.715&45.867&43.258&10.20&8, 9&3.86&$-1.599$\\
0847$-$120&0.566&13.11&LSF&20.29&1.66&44.949&43.790&43.644&41.030&9.07&3, 9&3.96&$-3.394$\\
0923$+$392&0.695&12.64&LSF&4.80&  &45.976&45.389&44.468&43.105&9.28&20, 6&3.63&$-2.005$\\
0945$+$408&1.249&12.34&LSF&7.37&1.54&45.858&45.124&44.763&43.028&8.95&3, 21&3.62&$-1.940$\\
0953$+$254&0.712&12.66&LSF&4.80&1.66&45.366&44.762&44.224&42.861&8.63&3, 21&2.47&$-1.982$\\
1156$+$295&0.729&13.07&LSF&34.29&1.52&45.623&44.333&44.255&41.185&8.89&5, 9&3.42&$-2.761$\\
1222$+$216&0.432&13.46&LSF&5.89&2.12&45.316&44.515&43.219&41.678&8.87&8, 21&2.48&$-2.469$\\
1226$+$023&0.158&13.85&LSF&19.88&1.60&45.733&44.608&43.957&41.361&8.69&23, 9&2.99&$-2.196$\\
1253$-$055&0.536&12.71&LSF&30.56&1.57&46.211&44.939&44.827&41.857&8.63&24, 9&4.64&$-1.805$\\
1324$+$224&1.400&12.48&LSF&25.09&1.49&46.167&45.005&44.780&41.981&9.24&3, 21&3.52&$-2.348$\\
1502$+$106&1.839&12.70&LSF&14.50&1.32&46.269&45.371&44.841&42.518&9.13&25, 9&3.45&$-1.873$\\
1510$-$089&0.360&13.04&LSF&20.69&1.40&45.294&44.242&43.558&40.926&8.65&5, 6&3.51&$-2.522$\\
1606$+$106&1.226&12.99&LSF&29.86&1.31&46.094&44.958&44.783&41.833&8.77&3, 10&3.74&$-1.926$\\
1611$+$343&1.401&12.26&LSF&16.72&1.58&46.035&44.983&45.326&42.880&9.08&3, 10&4.04&$-2.211$\\
1633$+$382&1.814&12.47&LSF&25.47&1.46&46.992&45.839&45.495&42.683&9.53&26, 21&3.97&$-1.805$\\
1637$+$574&0.751&12.80&LSF&16.20&1.82&45.573&44.436&44.214&41.795&8.69&3, 9&3.22&$-2.368$\\
1641$+$399&0.593&13.01&LSF&9.12&1.76&45.728&44.845&44.800&42.880&8.88&8, 10&4.37&$-2.149$\\
1725$+$044&0.293&13.46&LSF&4.09&1.58&44.545&44.019&43.132&41.909&8.07&3, 6&3.48&$-2.165$\\
1730$-$130&0.902&12.63&LSF&12.81&1.61&45.786&44.823&45.094&42.879&8.45&27, 9&5.01&$-1.741$\\
1739$+$522&1.375&13.12&LSF&31.75&  &46.778&45.487&45.057&42.054&9.09&28, 9&4.00&$-1.717$\\
1741$-$038&1.054&12.72&LSF&23.22&1.35&46.111&45.041&45.010&42.279&9.29&27, 29&5.15&$-2.363$\\
1828$+$487&0.692&12.99&LSF&6.51&1.32&45.731&45.102&44.689&43.062&8.59&30, 31&3.93&$-1.602$\\
1928$+$738&0.302&13.33&LSF&1.79&1.84&45.221&44.942&42.825&42.236&8.91&8, 6&2.38&$-2.082$\\
2121$+$053&1.941&12.63&LSF&17.79&1.95&46.338&45.109&45.487&42.987&9.33&3, 9&3.98&$-2.335$\\
2134$+$004&1.932&12.92&LSF&18.77&1.71&46.789&45.639&45.829&43.282&9.53&11, 21&3.97&$-2.005$\\
2145$+$067&0.990&12.53&LSF&19.22&1.52&46.404&45.326&44.846&42.278&8.87&8, 32&3.45&$-1.658$\\
2201$+$315&0.295&13.24&LSF&7.44&1.79&45.069&44.258&43.294&41.551&8.87&33, 6&2.86&$-2.726$\\
2227$-$088&1.562&12.68&LSF&18.52&  &47.137&46.046&45.236&42.700&8.96&28, 21&4.01&$-1,028$\\
2230$+$114&1.037&12.47&LSF&19.22&1.34&46.305&45.303&45.283&42.715&9.09&8, 9&4.31&$-1.901$\\
2234$+$282&0.795&12.96&LSF&6.62&1.52&45.479&44.789&44.472&42.830&8.44&3, 9&3.98&$-1.765$\\
2251$+$158&0.859&13.11&LSF&43.33&1.53&46.333&44.952&45.354&42.081&9.17&34, 6&3.92&$-2.332$\\
 \hline
0003$-$066&0.347&12.92&LBL&5.40&1.67&44.647&43.996&43.579&42.115&8.93&3, 9&4.34&$-3.048$\\
0059$+$581&0.664&12.67&LBL&17.86&  &45.277&44.146&44.154&41.651&9.01&35, 9&4.63&$-2.978$\\
0138$-$097&0.733&13.24&LBL&5.90&  &45.733&45.037&43.940&42.398&9.84&36, 37&3.08&$-2.917$\\
0208$-$512&0.999&12.94&LBL&3.27&1.72&46.288&45.821&43.799&42.770&9.12&38, 9&3.05&$-1.413$\\
0214$+$083&1.400&13.56&LBL&6.60&1.69&46.186&45.451&44.269&42.630&8.40&3, 9&2.78&$-1.063$\\
0235$+$164&0.940&13.03&LBL&10.78&1.82&45.800&44.830&44.353&42.288&9.07&39, 18&3.53&$-2.357$\\
0336$-$019&0.850&12.67&LBL&5.12&  &45.390&44.750&44.310&42.891&8.98&40, 6&3.31&$-2.344$\\
0422$+$004&0.310&13.67&LBL&18.37&  &44.847&43.705&43.364&40.835&8.76&35, 9&2.11&$-3.169$\\
0426$-$380&1.110&12.68&LBL&$6.18^*$&1.93&45.692&44.920&44.379&42.797&8.77&41, 9&3.49&$-1.964$\\
0430$+$289&0.970&12.31&LBL&$6.00^*$&1.42&45.050&44.422&43.830&42.273&8.25&3, 9&3.85&$-1.942$\\
0521$-$365&0.055&13.63&LBL&3.08&1.76&43.531&43.081&42.052&41.075&7.79&42, 9&2.97&$-2.823$\\
0723$-$008&0.128&12.90&LBL&1.79&1.56&44.280&44.065&41.684&41.179&8.00&8, 9&2.59&$-2.049$\\
0735$+$178&0.424&13.44&LBL&11.03&1.56&44.665&43.775&43.802&41.716&8.30&14, 18&2.54&$-2.639$\\
0749$+$540&0.200&12.56&LBL&5.50&1.79&43.656&42.968&42.631&41.150&8.94&3,  &3.80&$-4.086$\\
0754$+$100&0.266&12.89&LBL&8.10&1.59&44.846&44.062&43.080&41.263&8.53&3, 9&2.67&$-2.582$\\
0808$+$019&1.148&12.52&LBL&12.30&1.66&45.373&44.407&43.210&42.030&8.71&3, 9&2.76&$-2.417$\\
0814$+$425&0.530&13.01&LBL&6.33&1.76&44.530&43.793&43.757&42.154&8.01&14, 43&3.63&$-2.331$\\
0820$+$225&0.951&13.09&LBL&3.00&  &45.125&44.694&43.297&42.343&7.96&44, 9&2.37&$-1.380$\\
0823$+$033&0.506&13.04&LBL&29.41&1.73&45.045&43.709&43.865&40.928&8.55&3, 43&2.91&$-2.955$\\
0828$+$493&0.548&12.92&LBL&8.39&  &44.865&43.030&43.255&41.408&8.69&36, 45&3.82&$-2.773$\\
0829$+$046&0.174&13.70&LBL&5.28&1.50&44.056&43.453&42.437&40.991&8.46&3, 46&2.35&$-3.121$\\
0829$+$089&0.941&13.66&LBL&4.04&1.84&45.537&44.963&43.152&41.939&8.73&3, 21&2.72&$-1.881$\\
0851$+$202&0.306&13.24&LBL&14.76&1.96&44.907&43.754&43.420&41.082&8.79&47, 29&2.54&$-3.150$\\
0954$+$658&0.367&13.07&LBL&4.37&  &45.100&44.521&43.174&41.893&8.53&48, 15&3.54&$-2.123$\\
0958$+$294&0.558&12.97&LBL&5.76&1.76&45.083&44.384&42.958&41.438&7.78&3, 9&3.22&$-1.510$\\
1055$+$018&0.894&12.87&LBL&$8.56^*$&1.80&45.868&44.997&44.733&42.868&9.50&5, 9&4.07&$-2.617$\\
1057$-$797&0.581&12.72&LBL&$11.77^*$&1.90&45.264&44.229&44.162&42.020&8.64&5, 9&3.99&$-2.525$\\
1144$-$379&1.048&12.60&LBL&17.72&1.96&45.447&44.216&44.822&42.326&8.69&5, 9&4.03&$-2.588$\\
1308$+$326&0.996&12.72&LBL&27.73&1.60&45.985&44.734&44.752&41.866&8.81&5, 21&4.61&$-2.190$\\
1404$+$286&0.077&13.25&LBL&2.72&2.21&43.950&43.485&41.860&40.991&8.73&49, 7&2.75&$-3.359$\\
1413$+$135&0.247&12.82&LBL&13.18&1.59&44.100&43.133&43.004&40.764&7.88&49, 50&4.67&$-2.861$\\
1418$+$546&0.153&13.68&LBL&8.31&1.88&43.709&42.826&42.530&40.690&9.03&11, 6&2.97&$-4.318$\\
1501$+$481&0.345&13.12&LBL&1.96&  &43.649&43.385&41.306&40.722&7.97&51, 9&2.02&$-2.699$\\
1514$-$241&0.049&13.96&LBL&11.69&1.60&43.245&42.420&41.862&39.726&8.10&52, 6&2.81&$-3.794$\\
1519$-$273&1.294&12.72&LBL&22.33&2.25&45.713&44.251&45.072&42.375&8.80&11, 53&4.07&$-2.663$\\
1538$+$149&0.605&13.33&LBL&9.46&1.77&45.005&44.104&44.018&42.066&8.94&3, 29&3.49&$-2.950$\\
1622$-$253&0.786&12.62&LBL&7.15&0.72&45.704&45.214&44.237&42.528&7.70&14, 9&5.03&$-0.600$\\
1749$+$096&0.322&12.90&LBL&37.10&1.93&44.952&43.420&43.816&40.677&7.98&14, 9&4.14&$-2.674$\\
1803$+$784&0.684&13.25&LBL&8.51&1.50&45.579&44.804&44.297&42.437&8.94&5, 9&3.47&$-2.250$\\
1807$+$698&0.051&13.89&LBL&3.78&1.30&42.937&42.494&41.672&40.517&8.49&54, 46&2.31&$-4.110$\\
1823$+$568&0.663&13.16&LBL&4.07&1.61&45.503&44.972&43.924&42.704&9.26&26, 45&3.54&$-2.401$\\
2007$+$777&0.342&13.38&LBL&7.92&1.98&44.690&43.797&43.213&41.415&7.50&55, 45&3.20&$-1.816$\\
2029$+$121&1.215&12.56&LBL&8.94&1.53&45.376&44.573&44.647&42.744&8.31&3, 9&3.91&$-1.851$\\
2131$-$021&1.285&12.92&LBL&16.75&1.66&45.952&44.867&44.477&42.029&8.75&3, 9&3.69&$-1.997$\\
2150$+$173&0.871&13.36&LBL&6.60&1.12&45.040&44.461&44.118&42.479&8.61&3, 9&3.57&$-2.263$\\
2200$+$420&0.069&13.59&LBL&9.13&1.91&43.994&43.063&42.062&40.141&8.23&56, 6&1.99&$-3.281$\\
2209$+$236&1.125&11.92&LBL&21.19&  &45.474&44.276&44.194&41.542&8.69&57, 9&4.10&$-2.527$\\
2214$+$241&0.505&13.04&LBL&5.17&1.86&44.718&44.038&43.324&41.897&8.44&3, 9&3.35&$-2.516$\\
2223$-$052&1.404&12.96&LBL&19.11&  &47.011&45.853&45.575&43.012&8.81&28, 29&4.68&$-1.071$\\
2240$-$260&0.774&13.60&LBL&$6.08^*$&1.92&44.804&44.041&43.898&42.331&8.47&3, 9&3.52&$-2.543$\\
2254$+$074&0.190&12.79&LBL&7.11&  &43.666&42.896&42.370&40.667&8.62&NED, 6&2.66&$-3.838$\\
\hline 
0048$-$097&0.634&14.12&IBL&8.46&1.96&45.389&44.474&43.890&42.035&8.85&11, 9&2.70&$-2.490$\\
0118$-$272&0.559&14.15&IBL&9.81&2.33&44.771&43.671&43.636&41.653&9.54&3, 37&2.67&$-3.983$\\
0607$+$710&0.267&14.60&IBL&2.50&2.99&43.814&43.285&41.127&40.331&8.87&58, 43&1.45&$-3.699$\\
0716$+$714&0.310&14.17&IBL&10.03&2.72&45.170&43.928&43.589&41.586&8.10&59, 18&2.43&$-2.286$\\
1053$+$494&0.140&15.29&IBL&2.49&2.30&43.147&42.711&41.023&40.230&9.03&61, 43&1.87&$-4.433$\\
1055$+$567&0.143&14.75&IBL&3.55&2.53&43.959&43.312&41.691&40.591&8.33&3, 9&1.86&$-3.132$\\
1215$+$303&0.130&15.21&IBL&3.72&2.53&43.922&43.251&41.901&40.760&8.12&61, 6&1.83&$-2.983$\\
1219$+$285&0.102&14.65&IBL&8.69&2.79&43.854&42.668&42.096&40.218&8.55&62, 9&2.21&$-3.996$\\
1222$+$488&0.647&14.66&IBL&3.56&  &44.565&43.923&42.265&40.162&8.91&NED, 21&2.18&$-3.091$\\
1246$+$586&0.847&14.88&IBL&5.79&2.53&45.457&44.560&43.697&42.172&9.15&14, 43&2.22&$-2.704$\\
1424$+$240&0.160&15.29&IBL&4.23&2.35&44.150&43.450&41.946&40.693&8.79&14, 63&1.66&$-3.453$\\
1458$+$224&0.235&14.82&IBL&3.38&2.62&44.067&43.429&41.699&40.642&8.81&27, 9&1.99&$-3.495$\\
1532$+$302&0.064&14.27&IBL&1.66&  &42.840&42.588&40.367&39.927&8.27&64, 9&1.36&$-3.796$\\
1532$+$372&0.143&14.20&IBL&2.22&  &43.596&43.199&40.742&40.049&7.83&36, 9&1.81&$-2.745$\\
1749$+$701&0.770&14.10&IBL&12.20&2.01&45.510&44.420&44.394&42.221&9.90&3, 43&3.36&$-3.594$\\
1914$-$194&0.137&15.05&IBL&3.69&2.24&43.427&42.815&41.867&40.733&8.90&3, 65&2.37&$-4.199$\\
2005$-$489&0.071&15.30&IBL&2.50&2.31&44.353&43.914&41.447&40.651&9.03&66, 6&1.86&$-3.230$\\
2201$+$044&0.028&14.44&IBL&1.37&  &42.114&41.957&40.179&39.906&8.10&44, 6&2.08&$-4.256$\\
            \enddata
            
		\tablecomments{Col\,(1): Name. Col\,(2): Redshift. Col\,(3): The synchrotron-peak frequency of blazars (Hz). Col\,(4): The Type of blazars, LSF (ISF) indicate low (intermediate)-synchrotron-peaked FSRQ, LBL (IBL) indicate low (intermediate)-synchrotron-peaked BL Lac objects. Col\,(5): The 5\,GHz Doppler factor ("*" indicate that the 5\,GHz Doppler factors are estimated by us). Col\,(6): The power law photon index. Col\,(7) Logarithm of the observational 2--10\,keV X-ray luminosity ($\rm{erg}\cdot{\rm{s}^{-1}}$). Col\,(8) Logarithm of the intrinsic 2--10\,keV X-ray luminosity that are corrected from $L_{\rm{X,obs}}$ using the Doppler factor ($\rm{erg}\cdot{\rm{s}^{-1}}$). Col\,(9): Logarithm of the observational 5\,GHz core radio luminosity ($\rm{erg}\cdot{\rm{s}^{-1}}$). Col\,(10): Logarithm of the intrinsic 5\,GHz core radio luminosity that are corrected from $L_{\rm{R,obs}}$ using the Doppler factor($\rm{erg}\cdot{\rm{s}^{-1}}$). Col\,(11): Logarithm of black hole mass ($M_\odot$). Col\,(12): References for X-ray data and BH mass. Col\,(13): Logarithm of radio-loudness. Col\,(14): Eddington-ratio: $\lambda_{\rm{int}}=\log(L_{\rm{X,int}}/L_{\rm{Edd}})$.\\
    References: (1) \cite{Shinozaki2006spectral}; (2) \cite{Kawakatu2007}; (3) \cite{Paliya2017general}; (4) \cite{Vestergaard2009mass}; (5) \cite{Ghisellini2010general}; (6) \cite{Woo2002active}; (7) \cite{Rakshit2020spectral}; (8)
    \cite{Ricci2017bat}; (9) \cite{Paliya2021central}; (10) \cite{Shaw2012}; (11) \cite{Donato2005six}; (12) \cite{Liu2006jet}; (13) \cite{Acosta2010redshift}; (14) \cite{Giommi2012}; (15) \cite{Fan2004black}; (16) \cite{Ghisellini2011high}; (17) \cite{Palma2011multiwavelength}; (18) \cite{Zhang2024FP}; (19) \cite{Paliya2019detection}; (20) \cite{Ueda2005asca}; (21) \cite{Shen2011catalog}; (22) \cite{Abdo2010suzaku}; (23) \cite{Piconcelli2005xmm}; (24) \cite{Fang2005high}; (25) \cite{Cerruti2011suzaku}; (26) \cite{Giommi2021x}; (27) \cite{Malizia2016integral}; (28) \cite{Warwick2012xmm}; (29) \cite{Wang2004connection}; (30) \cite{Shi2005far}; (31) \cite{Kovalev2020transition}; (32) \cite{Liu2006jet}; (33) \cite{Ricci2014narrow}; (34) \cite{Ogle2011blazar}; (35) \cite{Saxton2008first}; (36) \cite{Dwelly2017spiders}; (37) \cite{Xu2009bl}; (38) \cite{Perlman2011deep}; (39) \cite{Raiteri2006x}; (40) \cite{Bhattacharya2013}; (41) \cite{Ghisellini2009jet}; (42) \cite{Boissay2016hard}; (43) \cite{Wu2009debeamed}; (44) \cite{Wang2016chandra}; (45) \cite{Wu2002supermassive}; (46) \cite{Woo2005black}; (47) \cite{Pal2020}; (48) \cite{Gonzalez2012x}; (49) \cite{Liao2020x}; (50) \cite{Readhead2021relativistic}; (51) \cite{Ueda2001asca}; (52) \cite{Kaufmann2013}; (53) \cite{Sbarrato2012relation}; (54) \cite{Torresi2018x}; (55) \cite{Sambruna2008}; (56) \cite{Raiteri2009webt}; (57) \cite{Giommi2007swift}; (58) \cite{Carrera2007xmm}; (59) \cite{Ferrero2006disentangling}; (60) \cite{Tavecchio2010tev}; (61) \cite{Aleksic2012discovery}; (62) \cite{Acciari2009}; (63) \cite{Padovani2022pks}; (64) \cite{Mateos2015revisiting}; (65) \cite{Carangelo2003optical}; (66) \cite{Acero2010pks}.}
	\end{deluxetable*}
\end{longrotatetable}

\end{document}